\documentclass{article}

\usepackage{amsmath}
\usepackage{pdfpages}
\usepackage{graphicx}
\usepackage{cite}
\usepackage{here}
\usepackage{amssymb}

\begin{document}

\author{Arndt Robert Finkelmann$^\dagger$, Hans Martin Senn$^\ddagger$\footnote{Corresponding author; e-mail: hans.senn@glasgow.ac.uk}, Markus Reiher$^\dagger$\footnote{Corresponding author; e-mail: markus.reiher@phys.chem.ethz.ch}    \\
        \small $^\dagger$ETH Zurich, Laboratorium f\"ur Physikalische Chemie, Vladimir-Prelog-Weg 2  \\
        \small 8093 Zurich, Switzerland\\[1ex]
        \small $^\ddagger$WestCHEM and School of Chemistry, University of Glasgow, Glasgow G12 8QQ, UK} 
\title{Hydrogen-Activation Mechanism of [Fe] Hydrogenase Revealed by Multi-Scale Modeling}
\maketitle

\begin{abstract}
When investigating the mode of hydrogen activation by [Fe] hydrogenases, not only the chemical reactivity at the active site is of importance but
also the large-scale conformational change between the so-called {\it open} and {\it closed} conformations, which leads to a special spatial 
arrangement of substrate and iron cofactor. To study H$_2$ activation, a complete model of the solvated and cofactor-bound enzyme in complex with the 
substrate methenyl-H$_4$MPT$^+$ was constructed. Both the {\it closed} and {\it open} conformations were simulated with classical molecular dynamics 
on the 100 ns time scale. Quantum-mechanics/molecular-mechanics calculations on snapshots then revealed the features of the active site that enable 
the facile H$_2$ cleavage. The hydroxyl group of the pyridinol ligand can easily be deprotonated. With the deprotonated hydroxyl group and the structural
arrangement in the {\it closed} conformation, H$_2$ coordinated to the Fe center is subject to an ionic and orbital push--pull effect
and can be rapidly cleaved with a concerted hydride transfer to methenyl-H$_4$MPT$^+$. An intermediary hydride species is not formed.
\end{abstract}

\newpage
\section{Introduction}
\label{sec:Intro}

\begin{figure}[h!]
 \centering
 \includegraphics[width=0.5\textwidth]{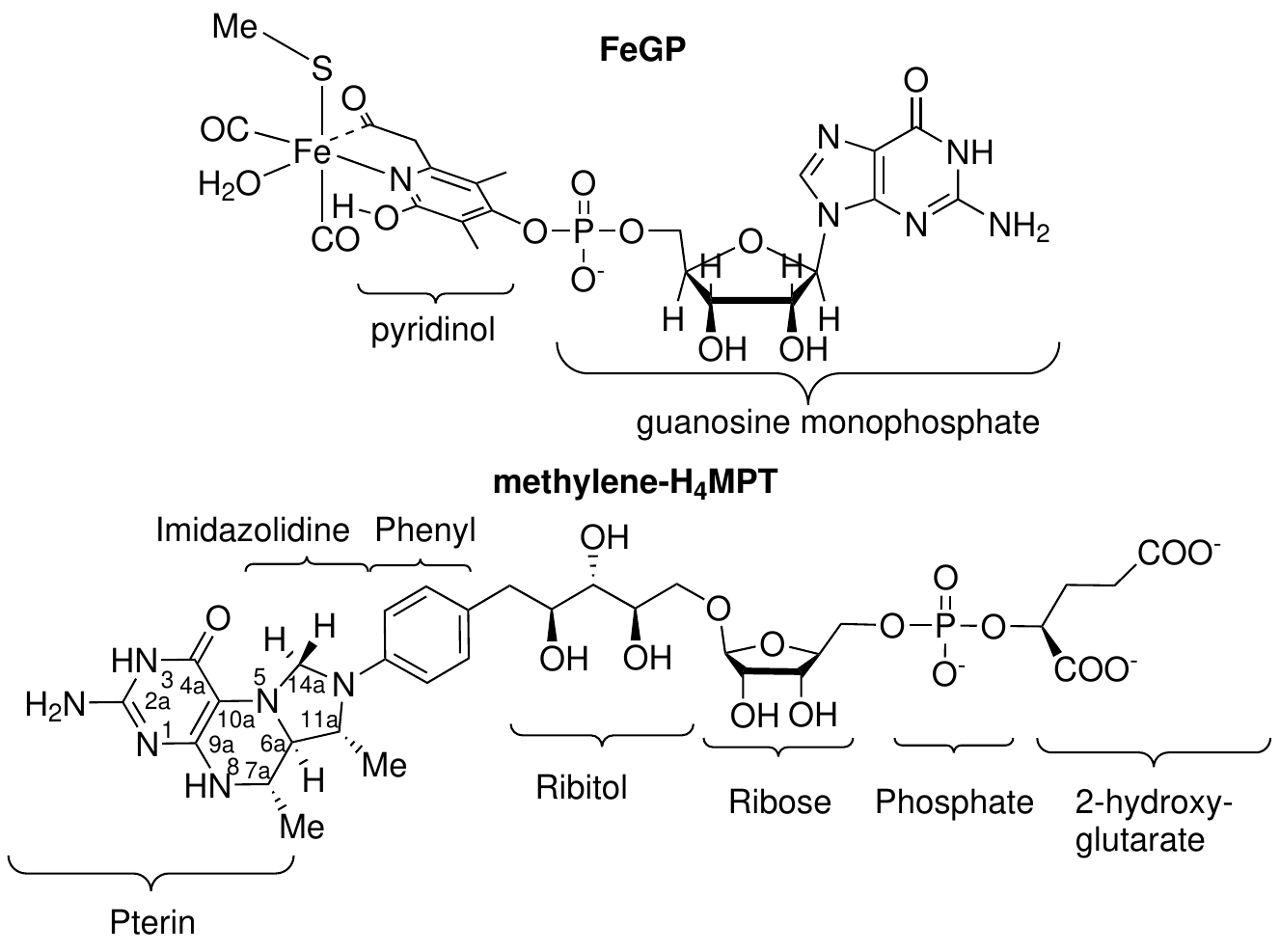}
 \caption{Top: Lewis structure of the FeGP cofactor. 
Bottom: Lewis structure of methylene-H$_4$MPT. The parts of methylene-H$_4$MPT are denoted according to their 
chemical building blocks, following Ref.\ \citenum{Hiromoto2009a}. Both molecules are depicted as parametrized for 
molecular-dynamics simulations (the cysteinate ligand is modeled by a methylthiolate).}
 \label{fig:FeGP_mpt_lewis}
\end{figure}

[Fe] hydrogenase \cite{Schwoerer1993,Klein1995,Hartmann1996,Zirngibl1992,Corr2011} , which features a mononuclear iron complex in the active site, 
differs in the mode of action compared to 
[NiFe] and [FeFe] hydrogenases \cite{Lubitz2014,Vincent2007,Fontecilla-Camps2007,Lubitz2007,DeLacey2007}.
In [FeFe] and [NiFe] hydrogenases, direct H$_2$ cleavage or formation is a redox process accomplished by oxidation 
state changes of the active Fe and Ni atoms, respectively \cite{Lubitz2014,Vincent2007,Fontecilla-Camps2007,Lubitz2007,DeLacey2007}. 
The reaction catalyzed by [Fe] hydrogenase,
\begin{equation*}
 \rm methenyl-H_4MPT^+ + H_2 \rightleftarrows methylene-H_4MPT + H^+,
\end{equation*}
is fundamentally different. No oxidation-state change of the active iron could be detected experimentally \cite{Lyon2004,Shima2005,Wang2008,Salomone-Stagni2010}. 
[Fe] hydrogenase requires the substrate methenyl-H$_4$MPT$^+$ (see Fig.~\ref{fig:FeGP_mpt_lewis}), 
which acts as the hydride acceptor. The question is therefore why the iron cofactor iron-guanylylpyridinol (FeGP) (Fig.~\ref{fig:FeGP_mpt_lewis})
is required for catalysis \cite{Lyon2004,Lyon2004a} even though it does not seem to be redox-active.

An accurate mechanistic description of the H$_2$ activation process must thus be able to account for the intriguing role of the metal cofactor. 
Yang and Hall were the first to investigate the mechanism computationally, using a truncated active-site model
in an electrostatic continuum \cite{Yang2009}. The first main step of the catalytic cycle is the heterolytic H$_2$ cleavage, with the
proton transferred to either the oxypyridine ligand (deprotonated pyridinol) or the cysteinate ligand, which need to be present in the deprotonated 
form. The second main step is hydride transfer to methenyl-H$_4$MPT$^+$ \cite{Yang2009}, which is the rate-limiting step \cite{Yang2009}. However, 
the theoretical description with density-functional-theory (DFT) methods is sensitive to the incorporation of empirical dispersion corrections 
and the energetics of all elementary reaction steps can be manipulated by first-shell ligand modifications \cite{Finkelmann2013a}.
Yang and Hall formulated the product of H$_2$ cleavage as a bound dihydrogen species 
with an elongated, polarized H--H bond, Fe(II)$\cdots$H$^{\delta-}$---H$^{\delta+}\cdots^-$O. This species is the 
resting state in their catalytic cycle \cite{Yang2009}. This intermediate could, however, also be described as a hydride complex 
\cite{Finkelmann2013a,Gubler2013}.

The difficulties fully to reconcile the experimental observations with that mechanism, which was derived based on a small model, point to the 
need for an extended theoretical treatment of the reaction. Any mechanistic proposals should be compatible with the following experimental data: 
(i) If the key intermediate was a stable hydride species, the electronic structure of the Fe atom would change. However, experimental 
and theoretical M\"ossbauer spectra indicate that no stable hydride or H$_2$-bound species exists under turnover conditions \cite{Shima2005,Gubler2013,Hedegard2014}. 
(ii) Model compounds that accurately mimic the first coordination sphere of the iron do not even bind H$_2$ \cite{Chen2011,Chen2011a,Chen2010}, hence, 
the protein environment is likely to be crucial. (iii) In model compounds, protonating the thiolate ligand 
leads to dissociation of the ligand \cite{Chen2012}. This indicates that also in the enzyme, the thiolate might not be a viable proton 
acceptor. (iv) Mutating histidine 14 to alanine reduces the catalytic activity to $1 \, \%$ of the 
wild-type level \cite{Shima2008,Hiromoto2009a}. The mechanism should thus explain why His14 is important for the catalytic activity.

\begin{figure}[h]
\centering
\includegraphics[width=0.5\textwidth]{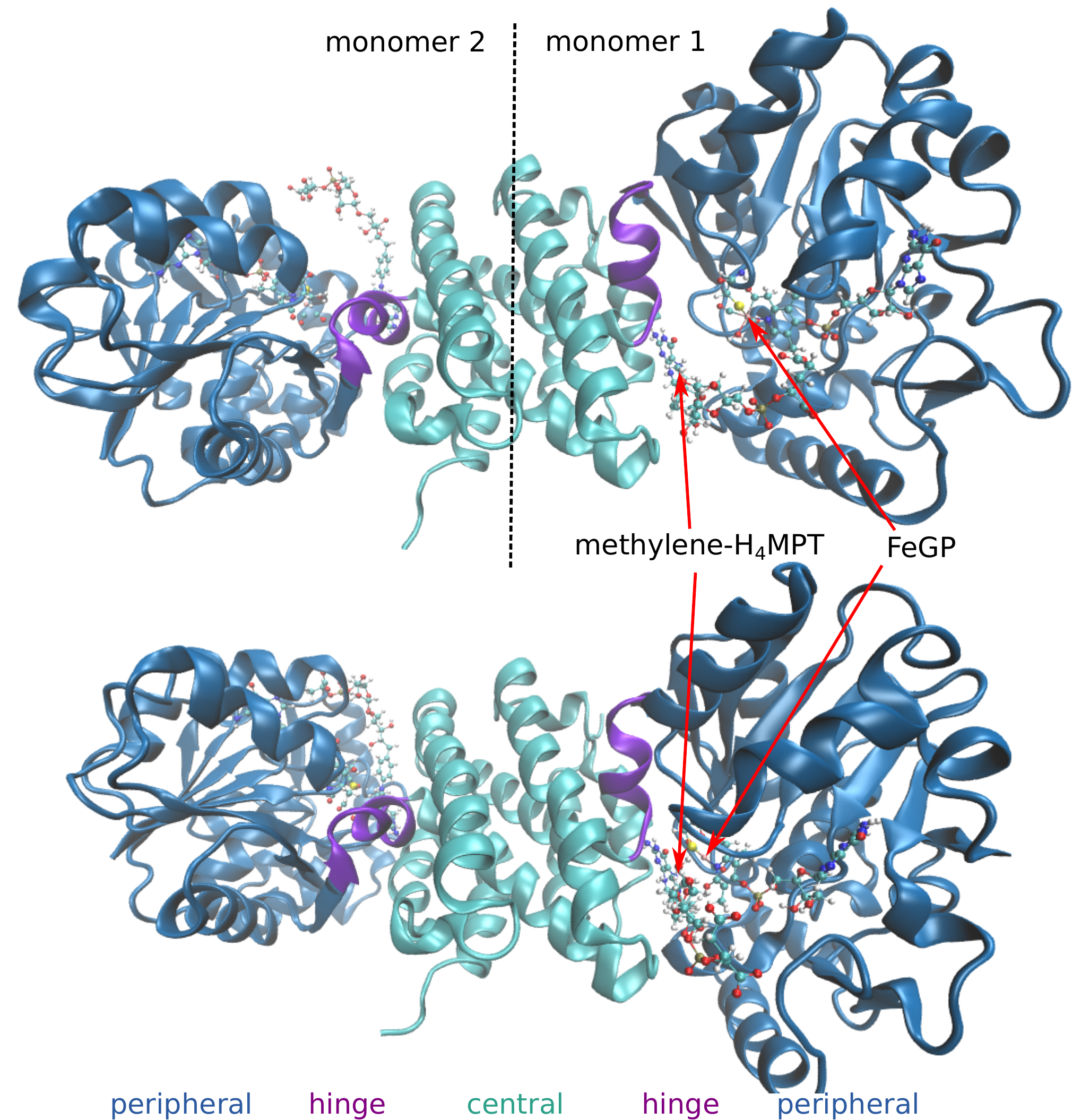}
\caption{Cartoon structure of our model of the substrate- and cofactor-bound protein dimer in the {\it open} conformation (top) and in the {\it closed} 
conformation (bottom). The red arrows point to the Fe-center and the hydride-accepting carbon atom of the substrate.}
\label{fig:QMMM_protein_open_closed}
\end{figure}

Already in 2009, Hiromoto \textit{et al.}\ suggested that a large-scale protein motion might play a role 
in catalysis \cite{Hiromoto2009a}. The dimeric protein has three subunits (see Fig.~\ref{fig:QMMM_protein_open_closed}): 
The central subunit, which is formed from the intertwined C-terminal domains of both monomers, and two identical peripheral subunits \cite{Pilak2006}. 
The peripheral subunits each harbor an FeGP cofactor \cite{Shima2008}; methenyl-H$_4$MPT$^+$ binds to the central subunit \cite{Hiromoto2009a}. 
The protein can adopt two conformational states, referred to as {\it open} and {\it closed}, respectively. In the 
{\it open} conformation, there is a cleft between the central and the peripheral subunits, as shown in Fig.~\ref{fig:QMMM_protein_open_closed} \cite{Shima2008,Hiromoto2009}.

Hiromoto \textit{et al.}\ proposed a mechanism where binding of methenyl-H$_4$MPT$^+$ to the {\it open} enzyme induces 
the transition to the {\it closed} conformation, in which methenyl-H$_4$MPT$^+$ and FeGP are arranged such 
that H$_2$ binding and cleavage can occur and methenyl-H$_4$MPT$^+$ is reduced to methylene-H$_4$MPT. Thus, the {\it closed} 
enzyme is the reactive conformation. Transition back to the {\it open} conformation and dissociation of the product 
methylene-H$_4$MPT closes the catalytic cycle. Hiromoto \textit{et al.}\ further postulated that the geometrical arrangement of FeGP and 
methenyl-H$_4$MPT$^+$ imposed by the protein in the {\it closed} conformation is necessary for catalysis to occur.

To model the reaction accurately, a method is required that incorporates the geometrical
constraints imposed by the protein and, if possible, also the electronic polarization exerted by the environment.
Combined quantum mechanics/molecular mechanics (QM/MM) fulfils both these requirements \cite{Warshel1976,Senn2009}. 
A further complication is that a crystal structure of the {\it closed} conformation is only available for the apoenzyme 
\cite{Pilak2006}, while the holoenzyme could only be crystallized in the {\it open} conformation \cite{Shima2008,Hiromoto2009}. 
A crystal structure of the enzyme in complex with the substrate is available only for the C176A mutant in the {\it open} 
conformation. Thus, suitable starting structures for the wild-type holoenzyme--substrate complex in the {\it open} and {\it closed} 
conformations must be devised first. 

To investigate the crucial H$_2$ cleavage step, we generated such starting structures of the enzyme--substrate complex in 
the {\it open} and {\it closed} conformations. In the subsequent molecular-dynamics (MD) simulations, the FeGP cofactor and 
methylene-H$_4$MPT (compare Fig.~\ref{fig:FeGP_mpt_lewis}) were parametrized with the General Amber Force Field (GAFF) \cite{Wang2004}.
We chose to simulate the enzyme--product complex because methylene-H$_4$MPT is more straightforward to parameterize with GAFF than 
methenyl-H$_4$MPT$^+$. According to the principle of microscopic reversibility \cite{Tolman1938}, this corresponds to the product 
structure directly after hydride transfer, and the sampled configurations are relevant for both reaction directions. The Fe atom, 
both CO ligands, the cysteinate S atom, the acyl CO of the pyridinol ligand, and the oxygen atom of the bound water were positionally 
restrained to avoid the need for Fe--L bonded parameters. As FeGP is strongly bound to the peripheral subunit, these restraints effectively 
lock the hinge motion that would interconvert {\it open} and {\it closed} conformations. However, this conformational change is likely to take
place on a time scale much longer than the sampling times used here.
The protein was described with the Amber ff03 force field \cite{Duan2003,Lee2004}. MD simulations were run with {\sc Gromacs} 4.5.5 
\cite{Berendsen1995,vanderSpoel2005,Hess2008,Pronk2013} ({\it open} conformation: $100 \rm \, ns$, {\it closed} conformation: $95 \rm \, ns$). 
Starting from snapshots of the MD trajectory, the H$_2$ splitting reaction was investigated by QM/MM calculations. These were performed with 
{\sc ChemShell} \cite{chemshell,Sherwood2003,Metz2014} interfaced to Turbomole \cite{Ahlrichs1989,Turbomole65} as QM back-end. QM calculations 
used the TPSS-D3 \cite{Tao2003,Grimme2010} DFT method with the def2-TZVP \cite{Weigend2005} basis set for iron and the def2-SVP 
\cite{Schaefer1992} basis set on all other atoms. The effect of a larger basis set was assessed for one reaction step 
and the differences in energies and structures were found to be negligible. Details on model construction and computational methods can be found 
in the electronic supporting information (ESI).

Herein, we use a full model of the dimeric enzyme in molecular-dynamics simulations and QM/MM calculations to address two central questions relating 
to the H$_2$-activation mechanism in [Fe] hydrogenase. These are the protonation state of the FeGP cofactor and the possible H$_2$-activation pathways 
in the {\it closed} conformation.

\section{Molecular dynamics simulations}
\label{sec:MD_analysis}

\subsection{{\it Open} conformation}
\label{sec:QMMM_MD_OPEN}
\begin{figure}[htb]
\centering
\includegraphics[width=0.5\textwidth]{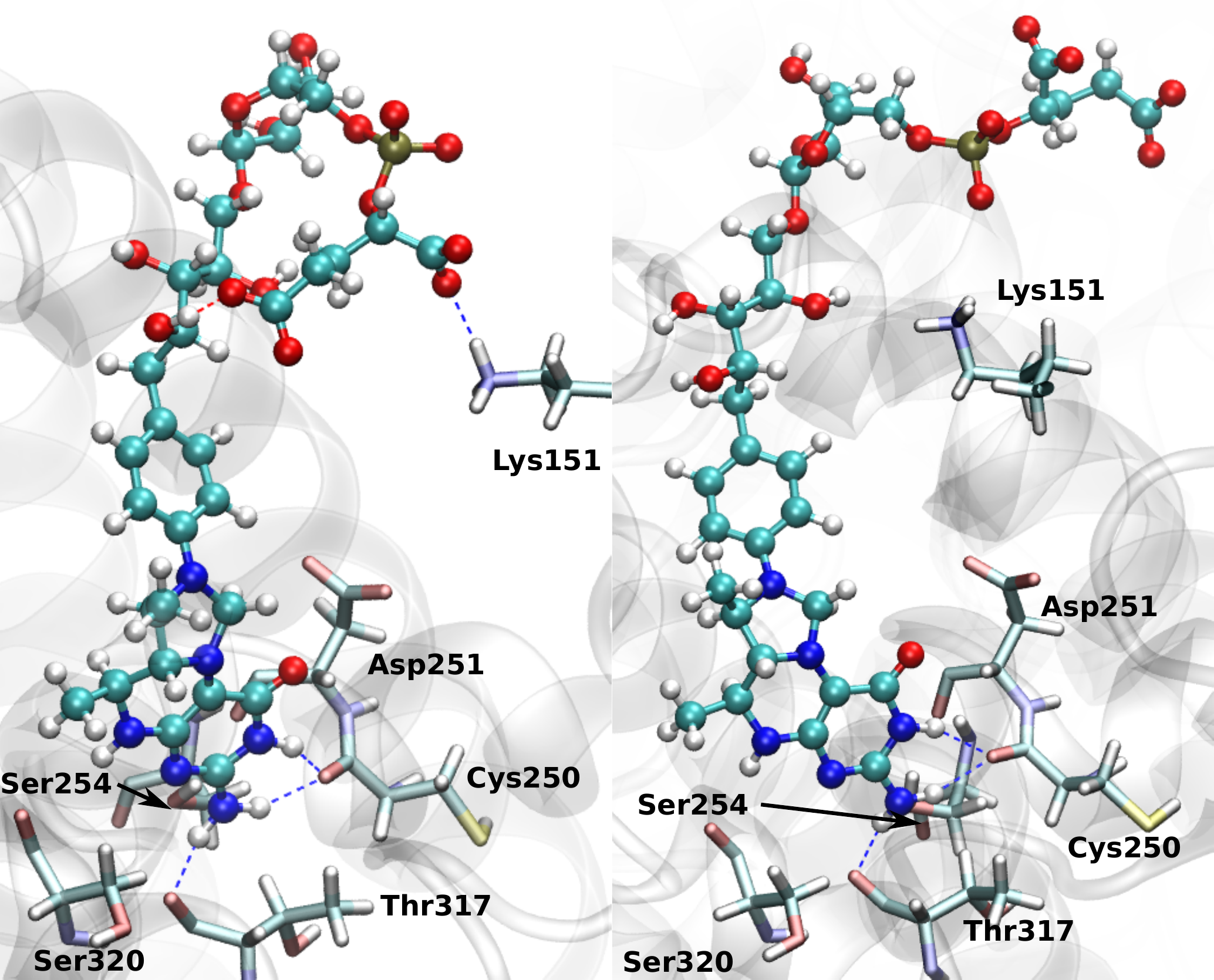}
 \caption{Representative snapshots of methylene-H$_4$MPT in the U-shaped conformation (left) and in an extended conformation (right).}
\label{fig:QMMM_h4mpt_two_conformations}
\end{figure}

The MD simulations of the dimer with both monomers in the {\it open} conformation yield insights into the dynamics around the cofactor
in this non-reactive conformation. The methylene-H$_4$MPT molecules (one bound to each monomer) stay attached to the central subunit 
of the protein throughout the simulation, mainly due to hydrogen-bonding interactions. Most of these interactions 
were already identified in the crystal structure \cite{Hiromoto2009a} and remained largely stable throughout the MD trajectory. 
Important hydrogen bonds are formed between the 2a-amino group of the pterin unit (see Fig.~\ref{fig:QMMM_h4mpt_two_conformations}) to the backbone carbonyls of 
Thr317 and Cys250. The carbonyl group of Cys250 also forms a hydrogen bond with the pterin N3--H.
The hydroxyl of Ser320 occasionally forms a hydrogen bond to the pterin 2a-amino group and the carbonyl of Ser320 with the pterin
N8--H. The hydroxyl group of Ser254 occasionally also engages in a hydrogen bond to the pterin 2a-amino group.
The tail of methylene-H$_4$MPT is highly flexible and mainly involved in hydrogen bonds to surrounding water molecules. One relatively stable hydrogen bond is formed between Lys151 and either of the glutarate carboxylates (but also to
the phosphate). The tail can adopt an extended conformation, and occasionally the glutarate carboxylates form hydrogen bonds with 
the distant residues Asn153, Lys154, Lys182 or Lys131. However, the predominant conformation of the tail is U-shaped, with the bend at the ribose.
 Snapshots from the MD simulations with methylene-H$_4$MPT in either conformation are shown in Fig.~
\ref{fig:QMMM_h4mpt_two_conformations}.

\begin{figure}[htb]
  \centering
 \includegraphics[width=0.5\textwidth]{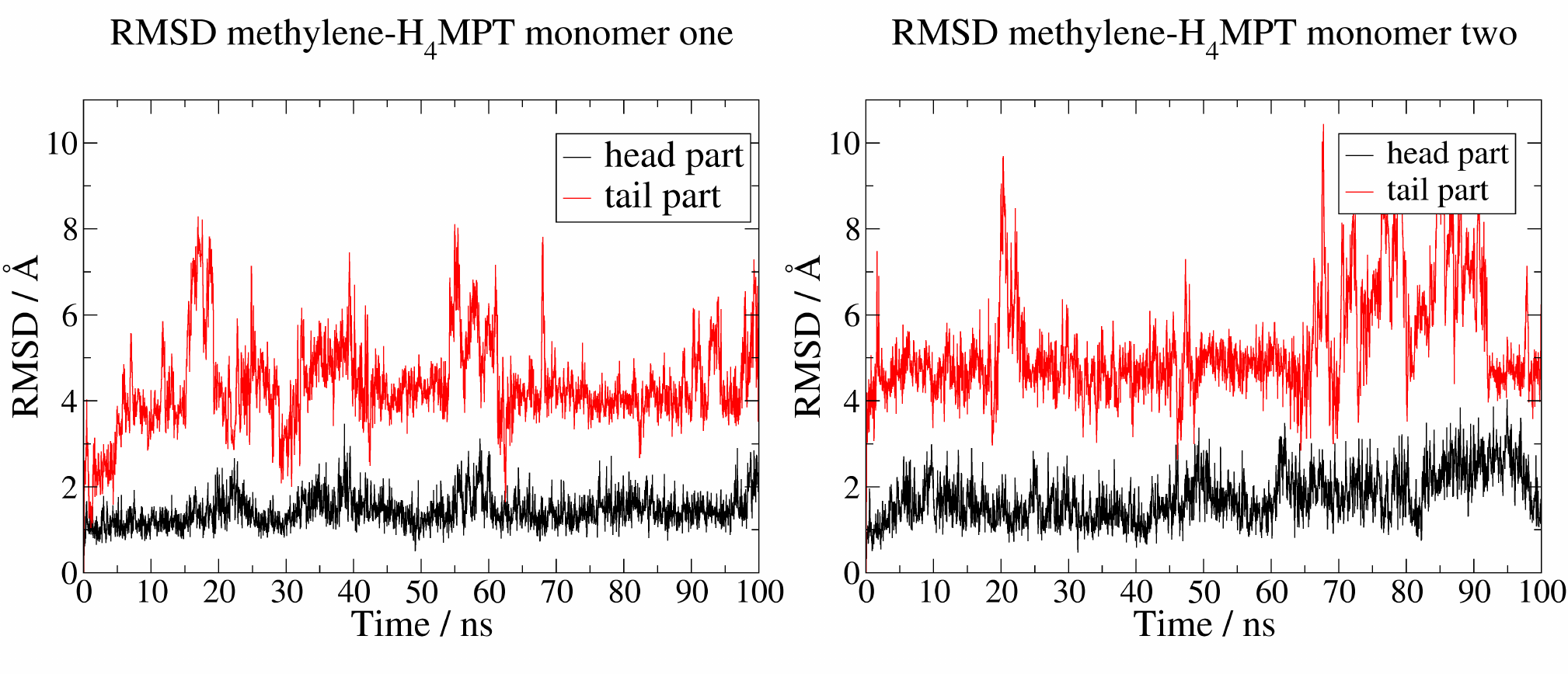}
 \caption{RMSD evolution of the head and tail parts of methylene-H$_4$MPT bound to monomer 1 (left panel) and monomer 2 (right panel) during the simulation of the {\it open} conformation. To calculate the RMSD, one 
frame was selected every $40 \rm \, ps$ and the protein backbone atoms of every structure were aligned.  }
\label{fig:QMMM_h4mpt_rmsd}
\end{figure}

The different conformational behavior of the head and tail parts of bound methylene-H$_4$MPT are mirrored in the RMSD (root-mean-square deviation) with 
respect to the starting structure.
The evolution of the RMSDs of the head and tail parts of both independent methylene-H$_4$MPT molecules over the whole trajectory are plotted in Fig.~\ref{fig:QMMM_h4mpt_rmsd}.
The RMSD of the head fluctuates between $1$ to $2 \, \textrm{\AA}$, so this part of the molecule remains essentially fixed. In contrast, the RMSD
of the conformationally flexible, very mobile tail is around $5 \, \textrm{\AA}$ (U-shaped tail), with values up to $10 \, \textrm{\AA}$ corresponding 
to the extended conformation.

\begin{figure}[htb]
 \centering
 \includegraphics[width=0.5\textwidth]{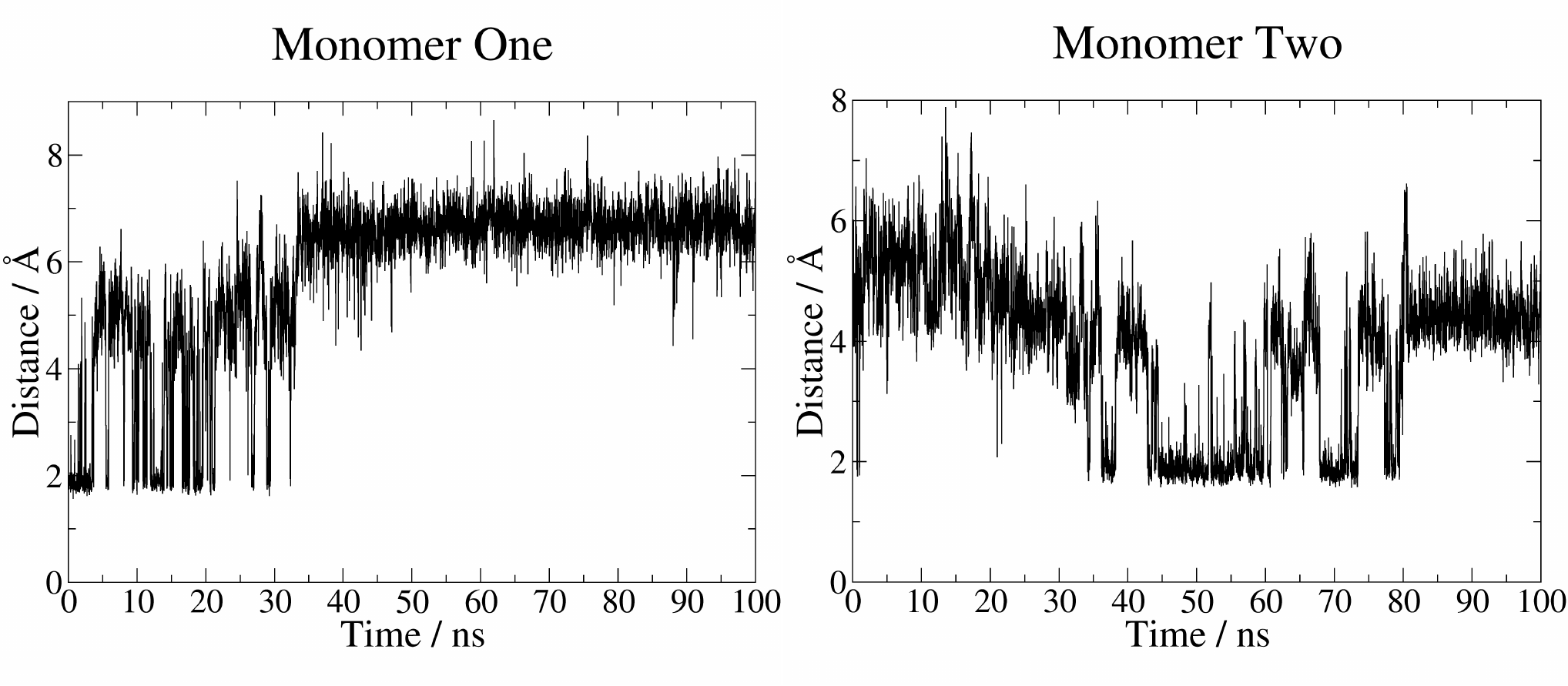}
 \caption{Distance between the proton of the pyridinol OH group and N$^{\epsilon}$ of His14 for both monomers during
the MD simulation of the {\it open} conformation.}
\label{fig:QMMM_open_oh_his_distances}
\end{figure}

In the {\it open} conformation, the Fe center is exposed to the solvent. The hydroxyl group of the pyridinol ligand mainly hydrogen-bonds 
to the water molecule coordinated to Fe (whose oxygen atom was positionally restrained). It can also form hydrogen bonds to bulk water molecules.
Interestingly, there is a relatively abundant conformation where the hydroxyl group forms a hydrogen bond to His14. His14 is known
to be crucial for high catalytic rates, since a H14A mutation reduces the turnover rate to $1 \, \% $ of the wild-type level \cite{Shima2008}.
The presence of this hydrogen-bonded conformation suggests that His14 may act as a base to deprotonate the pyridinol ligand, as previously suggested 
\cite{Shima2008,Hiromoto2009a}. The distance between the hydroxyl proton and N$^{\epsilon}$ of His14 
for both monomers is plotted in Fig.~\ref{fig:QMMM_open_oh_his_distances}. In monomer 1, this hydrogen bond 
is formed frequently at the beginning of the trajectory
but is no longer present beyond $36 \, \rm ns$, whereas in monomer 2, it was mainly observed later during the simulation (see Fig.~ \ref{fig:QMMM_open_oh_his_distances}). 

\subsection{{\it Closed} conformation}
\label{sec:QMMM_MD_CLOSED}
The simulation of the {\it closed} conformation samples the conformations of the protein in the state that
is believed to be reactive \cite{Hiromoto2009a}. The reduced substrate methylene-H$_4$MPT is bound in the active-site cleft with 
the hydride-accepting C14a atom in spatial proximity to the Fe atom of FeGP. The geometrical arrangement of methylene-H$_4$MPT
and the iron center is stable throughout the simulation. The mean distance between Fe and C14a increased from approximately $3.8 \, \textrm{\AA}$ 
($0$ to $30\, \rm ns$) to around $4.3 \, \textrm{\AA}$ ($30$ to $\, 90 \rm ns$) in monomer 1, while it remained constant at  approximately
$4.3 \, \textrm{\AA}$ throughout the simulation of monomer 2.
Notably, methylene-H$_4$MPT blocks a water channel identified in the crystal structure \cite{Hiromoto2009a}. 
However, a few water molecules are still able to enter the active site by passing along the cofactor already during the first
few nanoseconds of the MD simulation. 
This fast access of a few water molecules indicates that water is able to enter the active-site region through the cavity 
in the {\it closed} conformation. 

\begin{figure}[htb]
 \centering
 \includegraphics[width=0.5\textwidth]{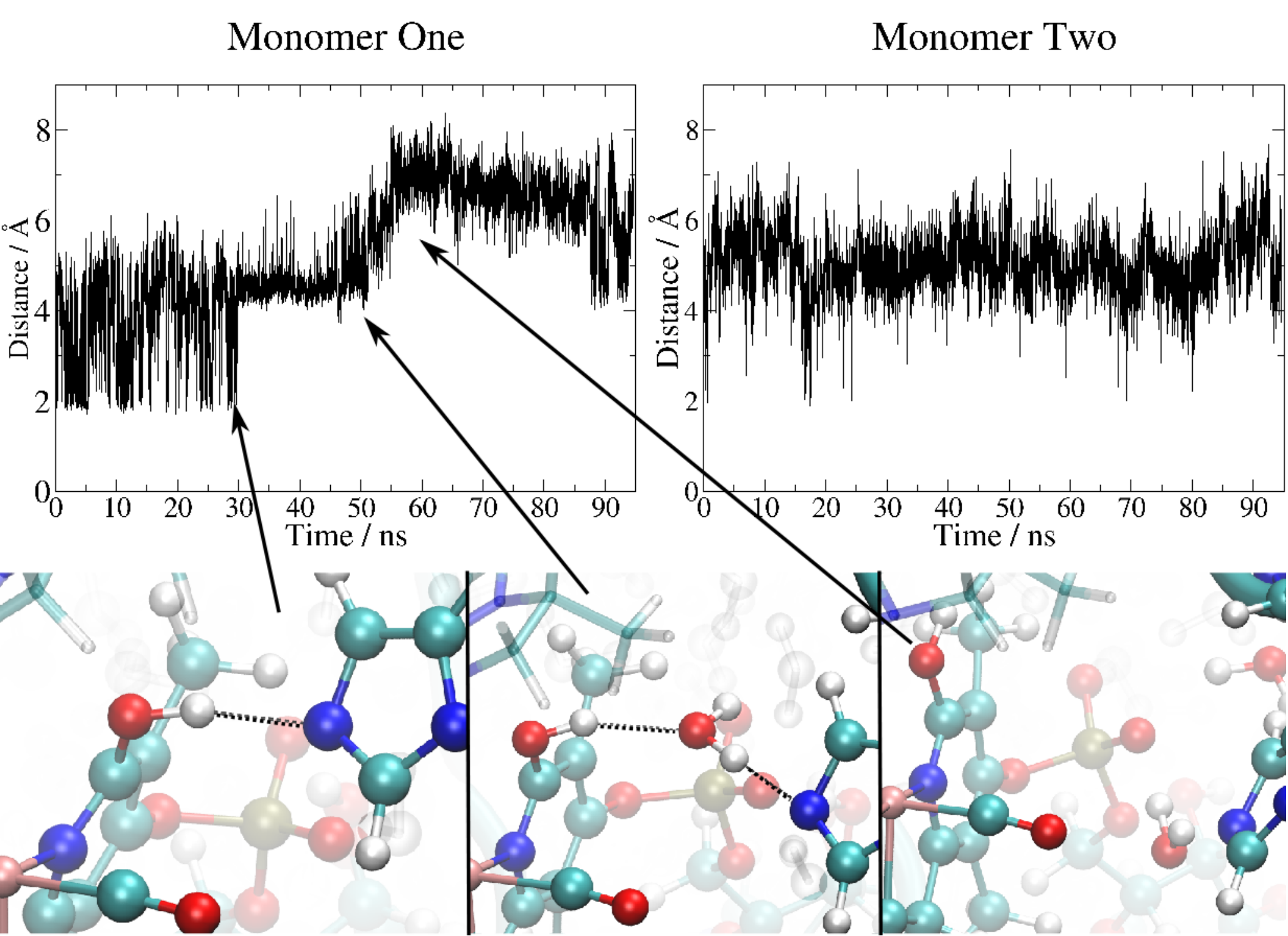}
 \caption{Top panels: Distance between the proton of the pyridinol OH group and N$^{\epsilon}$ of His14 for both monomers 
during the MD simulation of the {\it closed} conformation. Bottom panels:
Representative snapshots of the three hydrogen-bonding modes (direct hydrogen bond, one bridging water, no hydrogen bond).}
 \label{fig:QMMM_closed_oh_his_distances}
\end{figure}

The hydrogen bond between the pyridinol hydroxyl and His14, already observed 
in the {\it open} conformation, is formed in the {\it closed} conformation as well. The distance between the 
hydroxyl proton and N$^{\epsilon}$ of His14 for both monomers is plotted in Fig.~\ref{fig:QMMM_closed_oh_his_distances}.
In monomer 2, this hydrogen bond (OH--N$^{\epsilon}$ distance of  about $2 \, \textrm{\AA}$)
is frequently formed and broken over the course of the simulation. In monomer 1, the OH--N$^{\epsilon}$ distance plot shows three stages 
(see Fig.~\ref{fig:QMMM_closed_oh_his_distances}). In the first phase, up to $30 \rm \, ns$, a direct hydrogen bond is often formed. From 
$30$ to $51 \rm \, ns$, the OH--N$^{\epsilon}$ distance remains at around $4.5 \, \textrm{\AA}$. In this phase, a water 
molecule that entered the active site bridges the hydroxyl group and N$^{\epsilon}$ (OH--HOH--N$^{\epsilon}$). Thus, water-mediated 
proton transfer should still be possible. In the third phase, from $51$ to $88 \rm \, ns$, the hydrogen bond is lost and re-forms again 
only after $88 \rm \, ns$ with one bridging water molecule. Hence, in both monomers, deprotonation of the pyridinol with His14 as the base 
should be viable. The proton transfer may be mediated by a water molecule bridging between the hydroxyl and the proton-accepting N$^{\epsilon}$.

\section{QM/MM calculations}

\subsection{Protonation state of the guanylylpyridinol ligand}
\label{sec:QMMM_protonation_states}

\begin{figure}[htb]
 \centering
 \includegraphics[width=0.5\textwidth]{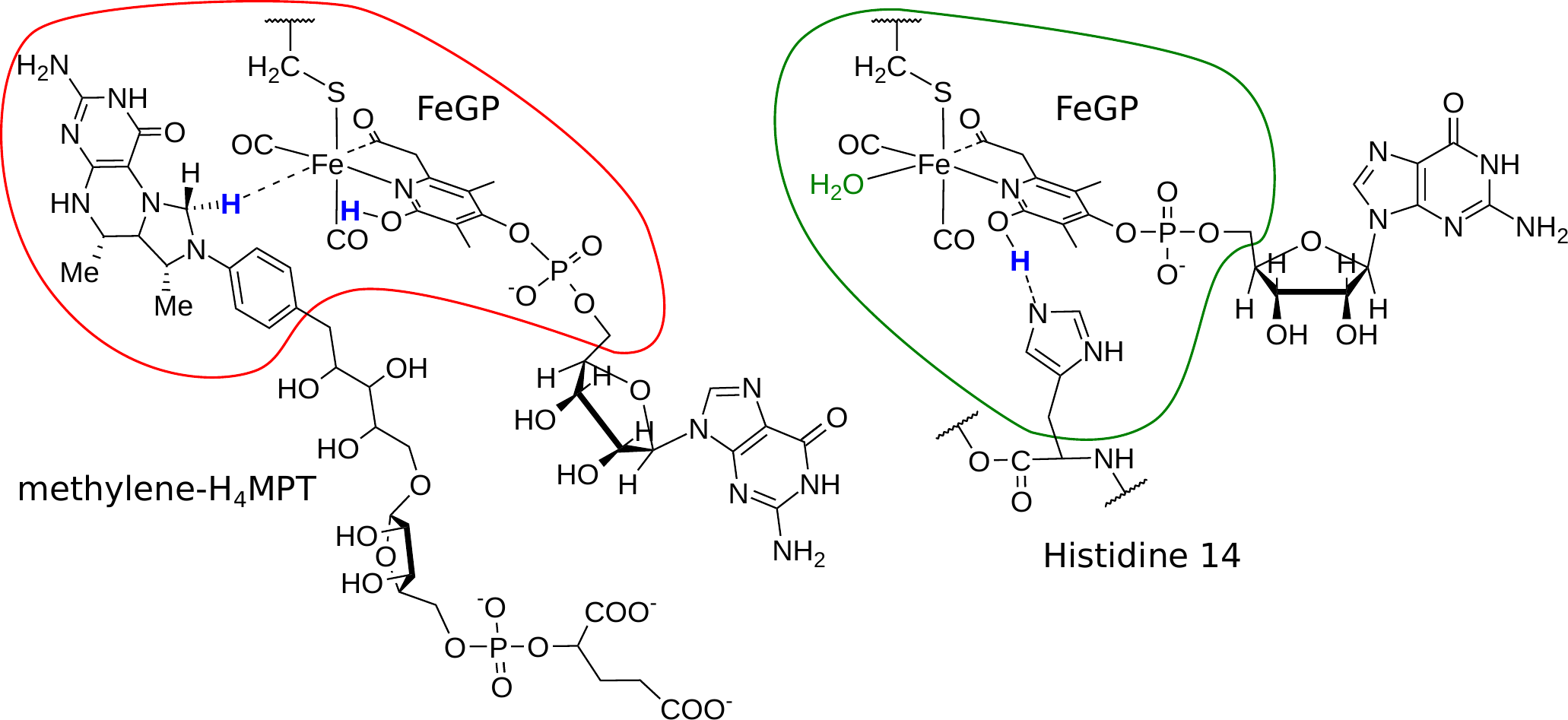}
 \caption{Left: Lewis representation of the active site of [Fe] hydrogenase with the reduced substrate methylene-H$_4$MPT. 
The QM region utilized to model hydrogen splitting is marked in red. Right: Lewis representation of the active site including His14. The 
QM region chosen to model the proton transfer from pyridinol to His14 (labeling according to PDB code 3H65 \cite{Hiromoto2009a})
is marked in green. The bound water molecule (green) is only present in the {\it open} conformation. Hydrogen atoms involved in the reactions 
are printed blue.}
\label{fig:QM_regions_lewis}
\end{figure}
\begin{figure}[htb]
 \centering
 \includegraphics[width=0.5\textwidth]{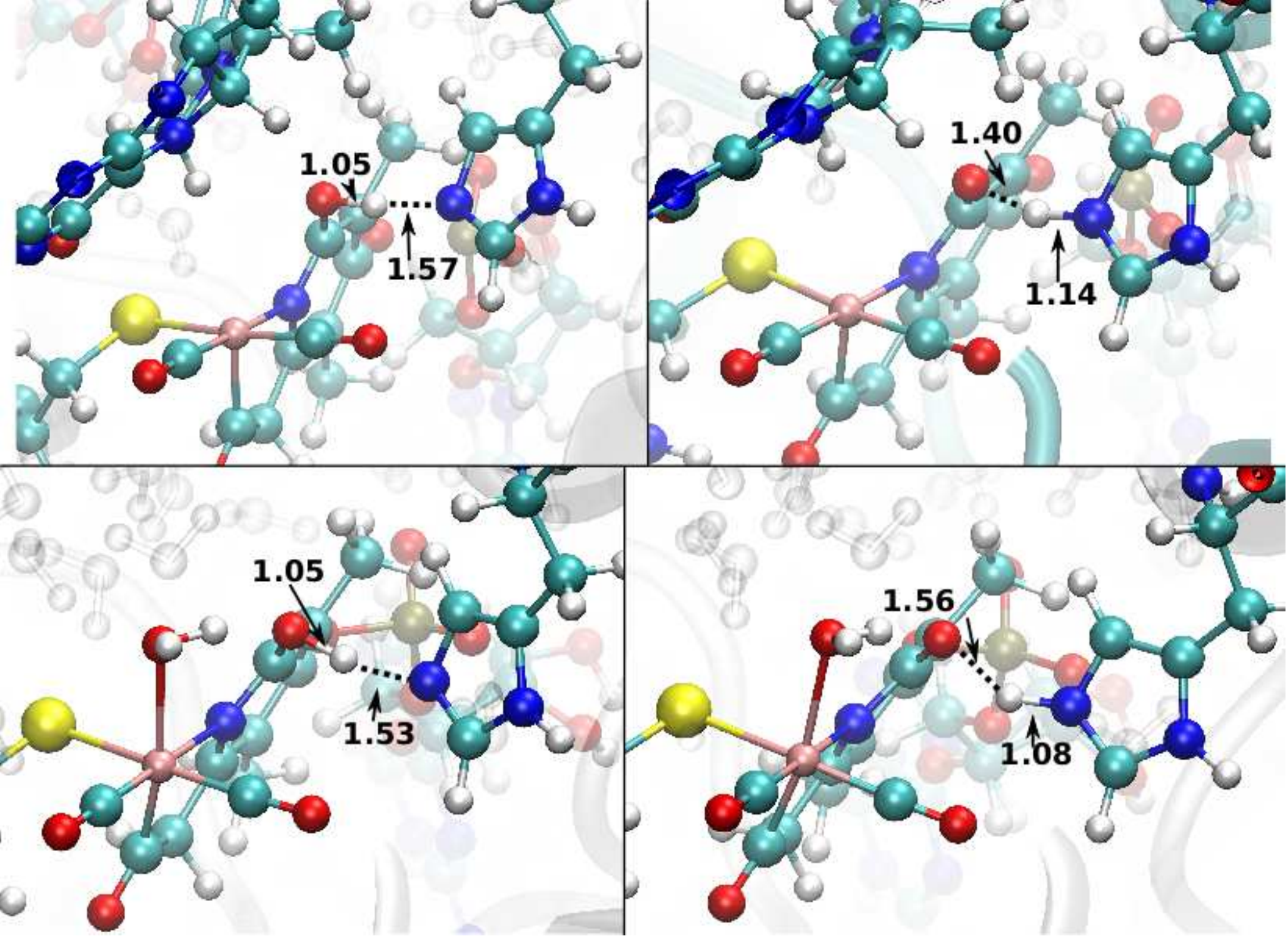}
 \caption{QM/MM-optimized reactant (left column) and product (right column) structures for the proton transfer from pyridinol OH to His14.
Top row: {\it closed} conformation; bottom row: {\it open} conformation with Fe-bound water. Water molecules in the active site
are shown as ``ghost atoms''; selected distances are given in $\textrm{\AA}$.}
\label{fig:QMMM_htransfer_reactants_products}
\end{figure}

To investigate possible H$_2$ activation mechanisms, we first need to clarify the protonation state of the active site. The experimentally verified importance of 
His14 for a high turnover rate \cite{Shima2008,Hiromoto2009a} and the hydrogen bond between His14 and the pyridinol OH observed in the MD simulations point to a 
crucial role of His14 as the base in the proton transfer pathway. Deprotonation of the hydroxyl 
group results in a potent proton acceptor (oxypyridine) for heterolytic H$_2$ cleavage. To investigate the energetics of pyridinol deprotonation, we chose two 
representative MD snapshots that feature the OH--His14 hydrogen bond: One snapshot from the {\it open} conformation (at $10.78 \, \rm ns$) and one from the {\it closed} 
conformation (at $13.2\, \rm ns$). Both snapshots were prepared for QM/MM optimization, \textit{i.e.}, the full protein
plus a water shell around one of the active sites was extracted (see ESI for details). The QM region contained the FeGP cofactor up to the phosphate linker 
(with an Fe-bound water in the {\it open} conformation) and the His14 side chain; see Fig.~\ref{fig:QM_regions_lewis}.
The optimized structures of the pyridinol/His and oxypridine/HisH$^+$ forms are presented in Fig.~\ref{fig:QMMM_htransfer_reactants_products}.

In the {\it closed} conformation, the proton transfer is endothermic by $+2.3 \rm \, kcal/mol$. From a potential-energy surface (PES) 
scan along the proton-transfer coordinate (defined as the difference between O--H and H--N$^{\epsilon}$ bond lengths), we estimate an upper bound 
for the proton-transfer barrier of  $+4.6 \rm \, kcal/mol$ ($+2.3 \rm \, kcal/mol$ for the back reaction). Thus, proton transfer between the pyridinol 
OH and His14 is facile, with the OH/His form being favoured. Considering that His14 is connected to the bulk solvent through a proton-transfer chain, we conclude 
that the less favoured oxypyridine (O$^-$/HisH$^+$) form is still present in significant amounts under equilibrium conditions.

In the {\it open} conformation, the proton transfer is thermoneutral ($ \Delta E = 0.0 \rm \, kcal/mol$).
The oxypyridine form is thus equally likely. Although we did not calculate the reaction barrier for this case, it is reasonable to assume that it will be 
similar to the barrier in the {\it closed} conformation as the proton transfer reactions are the same in both cases, except for a slight change in the environment. 
The stabilization of the oxypyridine form in the {\it open} conformation compared to the {\it closed} conformation arises because the water molecule coordinated 
to iron can form a hydrogen bond to the oxypyridine oxygen, stabilizing the anion. Note that the active site in the 
{\it open} conformation is exposed to the bulk solvent and thus filled with water (see Fig.~\ref{fig:QMMM_htransfer_reactants_products}). 
For the H$_2$ activation to proceed, the bound water must be displaced by H$_2$. Based on all available data, we cannot assess 
with any certainty if this happens while the enzyme is in the {\it open} or the {\it closed} conformation. As we have found that the 
active site is still accessible to water in the {\it closed} conformation (see Sect.\ \ref{sec:QMMM_MD_CLOSED}), it is certainly possible 
for H$_2$ to enter the active site only after the {\it closed} conformation has formed. What is clear, however, is that the prevailing 
protonation state of the pyridinol/His14 pair will critically depend on the external pH.

\subsection{H$_2$ activation}
To investigate hydrogen cleavage and hydride transfer to methenyl-H$_4$MPT$^+$, we chose two representative snapshots from the {\it closed} 
conformation: one with a short Fe--C14a distance of $3.7 \rm \, \mathrm{\AA}$ (at $11 \rm \, ns$) and one with a longer distance of 
$4.3 \rm \, \mathrm{\AA}$ (at $56.5 \rm \, ns$). Because the hydride is transferred to C14a of methenyl-H$_4$MPT$^+$, one might expect the 
reaction to be facilitated by a short Fe--C14a distance, and our discussion thus focuses first on the former snapshot. The QM region included 
again FeGP up to the phosphate linker, together with the chemically relevant part of the substrate and H$_2$ (see Fig.~\ref{fig:QM_regions_lewis}).
In the selected snapshot, the pyridinol-OH--His14 hydrogen bond is not present. Note that His14, in the neutral form, is in the MM region and does not 
directly participate in the reaction. Considering that the pyridinol--His14 hydrogen bond is frequently formed and broken (see Fig.~\ref{fig:QMMM_closed_oh_his_distances}) 
and that the proton transfer is kinetically facile (see Sect.\ \ref{sec:QMMM_protonation_states}), this choice of setup sustains two scenarios: 
(i) The pyridinol ligand has been deprotonated {\it via} His14, and the proton removed from the active site through the proton-transfer chain, leaving 
behind oxypyridine and neutral His14. (ii) The pyridinol ligand remains neutral, without hydrogen-bonding to (also neutral) His14.
\begin{figure*}[htb]
 \centering
 \includegraphics[width=\textwidth]{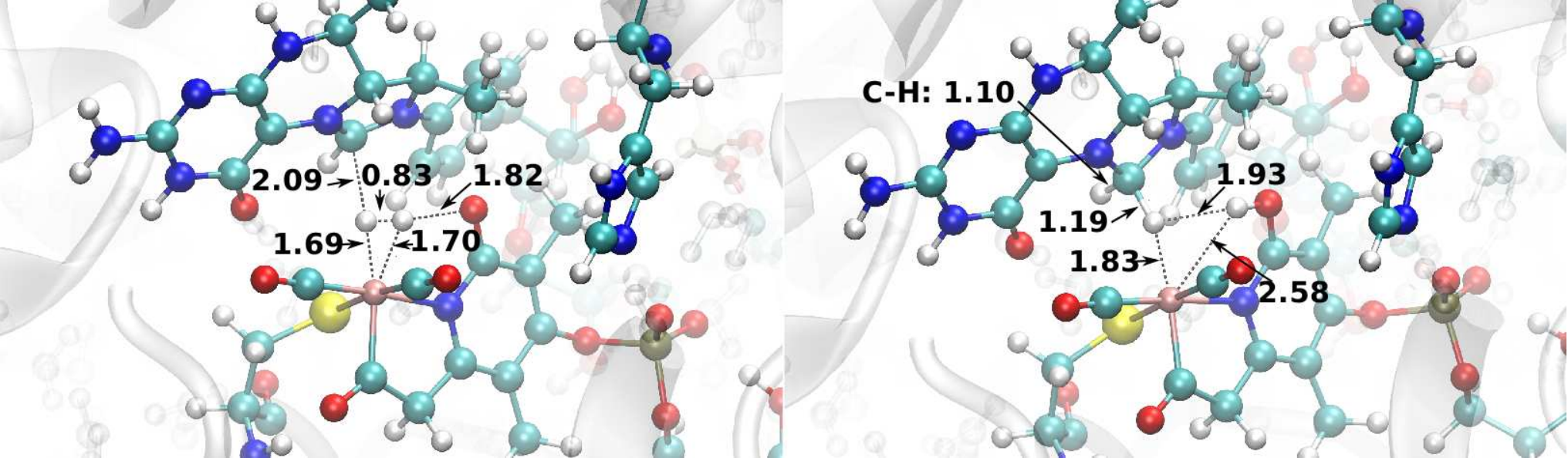}
 \caption{QM/MM-optimized reactant (left) and product (right) structures of the H$_2$ cleavage reaction for the scenario with oxypyridine ligand. 
Distances are given in $\textrm{\AA}$.}
\label{fig:QMMM_hydrogen_splitting}
\end{figure*}

\subsubsection{H$_2$ activation \textit{via} oxypyridine.~~}
For the scenario with oxypyridine, we studied several possible hydrogen coordination modes to the open coordination site at the 
Fe center: end-on and two rotamers of side-on coordination. All initial structures converged to a side-on-coordinated hydrogen molecule, which is thus the
reactant for the hydrogen cleavage reaction. The structure is shown in Fig.~\ref{fig:QMMM_hydrogen_splitting}. 
The coordinated H$_2$ is activated, its bond being elongated to $0.83 \, \textrm{\AA}$ (from $0.74 \, \textrm{\AA}$ in free H$_2$).
There is only one reasonable pathway to cleave H$_2$ in this configuration: In a concerted heterolytic cleavage step, the proton
is transferred to the oxypyridine oxygen and the hydride to C14a of methenyl-H$_4$MPT$^+$. This reaction is exothermic by $-18.7 \rm \, kcal/mol$. 
A PES scan (along the difference of O--H and H--H bond lengths) provided an upper bound for the barrier of about $+1.0 \rm \, kcal/mol$. 
Despite various attempts, we were unable to locate a stable minimum on the PES that would correspond to an iron hydride species. Hence, 
we find that the iron is involved in H$_2$ binding and activation, but does not bind a hydride species.
The H$_2$ cleavage mechanism we have identified here thus complies with the first of the requirements formulated in Sect.\ \ref{sec:Intro}.

\begin{figure}
 \centering
 \includegraphics[width=0.5\textwidth]{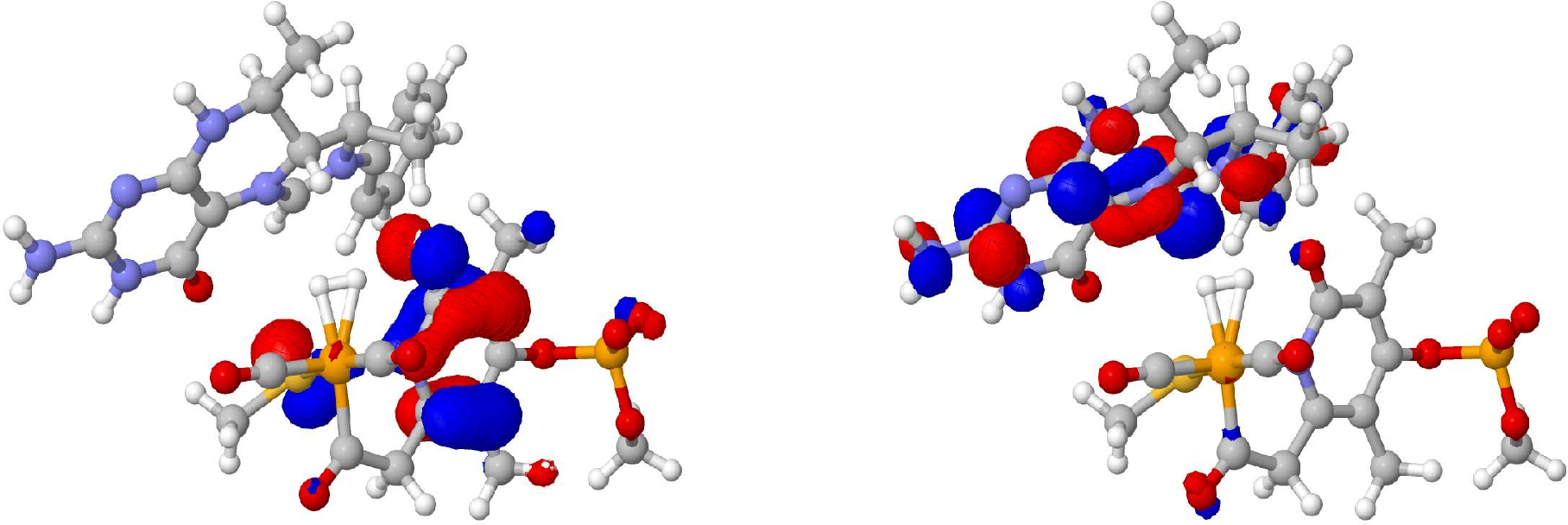}
 \caption{HOMO$-2$ (right) and LUMO (left) of the H$_2$ adduct of the oxypyridine form of the FeGP cofactor. }
\label{fig:orbitals}
\end{figure}

In the reactant complex, the coordinated H$_2$ is subjected
to an electronic push--pull effect from the negatively charged oxypyridine oxygen and the positively charged carbocation of methenyl-H$_4$MPT$^+$.
This is reflected in the relevant frontier orbitals (Fig.~\ref{fig:orbitals}): 
The LUMO has a strong contribution from the p$_{z}$ orbital on 
C14a, which is oriented perpendicular to the ring plane. The HOMO$-2$ is delocalized over the oxypyridine ring and the thiolate S atom, with a 
strong contribution from the oxypyridine oxygen. (Note that HOMO and HOMO$-1$ are strongly localized on the phosphate linker and thus do not 
contribute to the reactivity at the iron center).

In the optimized product structure resulting directly from the PES scan, the O--H bond of the pyridinol hydroxyl points towards the empty 
coordination site of the iron center (Fig.~\ref{fig:QMMM_hydrogen_splitting}). The newly formed C14a--H bond is pointing towards
the Fe atom. It is slightly elongated ($1.19 \, \textrm{\AA}$ compared to $1.10 \, \textrm{\AA}$ for the other C14a--H bond), indicating a weak 
interaction with the iron center. Another conformer, where the O--H bond has turned away from 
the iron, was found to be $2.3 \rm \, kcal/mol$ more stable.

Similar results were obtained for the second snapshot, where the substrate was positioned slightly further away from the iron center 
(4.3 \textit{vs.}\ $3.7 \rm \, \mathrm{\AA}$). The cleavage reaction in that case is even more exothermic (by $-26.3 \, \rm kcal/mol$, compared to 
$-18.7 \, \rm kcal/mol$ for the first snapshot), and again no iron hydride species could be optimized. We thus find that fluctuations 
of the substrate position of the order as they were observed in the MD simulations have only a minor effect on the reactivity of [Fe] 
hydrogenase in the H$_2$ cleavage step.

\subsubsection{H$_2$ activation \textit{via} thiolate.~~} 

\begin{figure}[htb]
 \centering
 \includegraphics[width=0.5\textwidth]{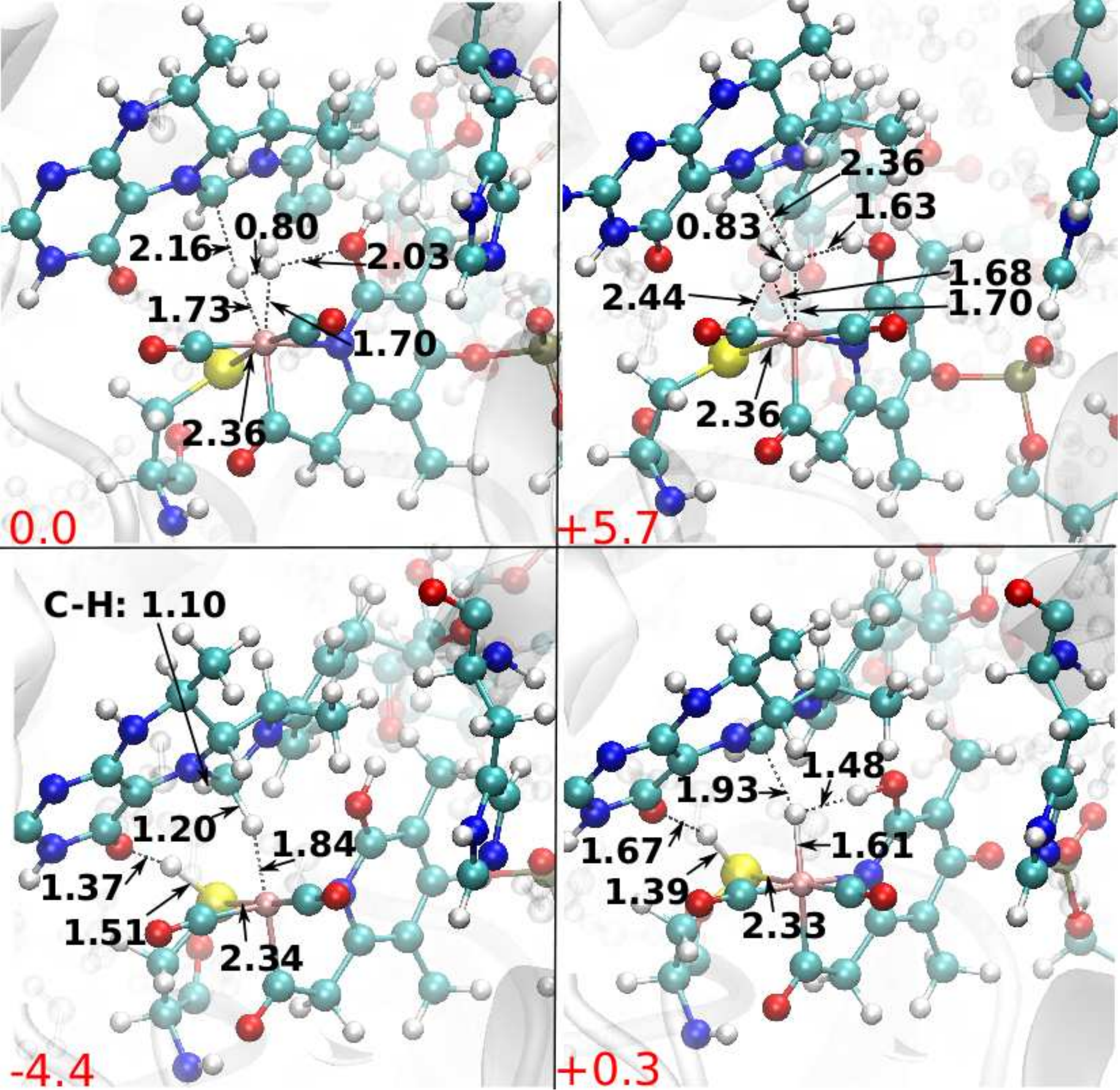}
 \caption{Top row: Structures of the H$_2$ adduct for the second scenario with neutral pyridinol; the pyridinol OH can be oriented
away from Fe (top left) or towards Fe (top right). Bottom row: Products of H$_2$ cleavage, with the proton transferred to the thiolate;  with the hydroxyl 
oriented away from Fe (bottom left) and towards Fe (bottom right). Distances are given in $\textrm{\AA}$; relative energies with respect to the favoured adduct are indicated in red in $\rm kcal/mol$ .}
\label{fig:QMMM_hydrogen_splitting2}
\end{figure}

In the second scenario, we consider the cleavage reaction with a neutral pyridinol ligand. There are two relevant reactant conformers, which
differ in the orientation of the pyridinol O--H bond (Fig.~\ref{fig:QMMM_hydrogen_splitting2}). In the following, we quote energies relative 
to the favoured conformer with the O--H pointing away from the iron. The second conformer, where the O--H is pointing towards the coordinated 
H$_2$, is $5.7 \rm \, kcal/mol$ less stable. Direct H$_2$ splitting in the favoured conformer, with the pyridinol OH acting as the proton acceptor, 
is not possible: Product-like starting structures, with one hydrogen atom already transferred to C14a of methenyl-H$_4$MPT$^+$, are not stable 
minima but converged back to reactant structures during optimization. For the second conformer, this pathway is precluded in the first place by 
the orientation of the pyridinol OH bond.

However, we find for both reactant conformers that hydrogen cleavage can occur with the 
thiolate ligand, rather pyridinol OH, as the proton acceptor. When the OH group is oriented away from
the Fe center, the resulting iron hydride structure is not stable but the hydride is directly transferred to methenyl-H$_4$MPT$^+$ to form the product. 
This reaction is exothermic by $-4.4 \rm \, kcal/mol$, significantly less so than hydride cleavage to oxypyridine-O$^-$.
For the reactant conformer with the OH bond oriented towards the Fe center, a stable iron hydride intermediate could indeed 
be located (Fig.~\ref{fig:QMMM_hydrogen_splitting2}). It is only $+0.3 \rm \, kcal/mol$ higher in energy than the favoured reactant conformer. 
The hydride Fe--H bond length is $1.61 \, \textrm{\AA}$, in excellent agreement with Fe--H bonds in comparable hydride complexes optimized {\it in vacuo} 
\cite{Gubler2013} ($1.60\, \textrm{\AA} $). 
OH rotation, which is likely to have a low activation barrier, triggers the transfer of the hydride from iron to methenyl-H$_4$MPT$^+$, yielding the 
same product as direct H$_2$ splitting from the preferred conformer (see Fig.~\ref{fig:QMMM_hydrogen_splitting2}). The thiolate is thus able to act as 
the base, which may provide an explanation for the $1 \, \%$ remaining activity of the H14A mutation 
\cite{Shima2008,Hiromoto2009a}.

Remarkably, the Fe--SH bond in the product ($2.34\, \textrm{\AA} $) is even slightly shorter than the Fe--S$^-$ bond in the reactant ($2.36\, \textrm{\AA} $). 
In the product, the thiol proton forms a short hydrogen bond ($1.37\, \textrm{\AA}$) to the pterin carbonyl group of 
methylene-H$_4$MPT (see Fig.~\ref{fig:QMMM_hydrogen_splitting2}), which is a very good hydrogen-bond acceptor because of its conjugation with the guanidine 
moiety in the pterin ring. This in turn weakens the S--H bond (elongated to $1.51\, \textrm{\AA}$, compared to $1.39\, \textrm{\AA}$ in the hydride intermediate), 
which may be described as ``partial deprotonation" of the thiol. The formal thiol ligand in the product is similar in character to a thiolate in terms of its 
interaction with the metal center. The thiol--pterin hydrogen bond thus makes the thiolate a better proton acceptor in the H$_2$ splitting step, stabilizes the 
thiol product, and also makes the thiol a better ligand, preventing it from dissociating like in model complexes. The hydrogen bond is enabled by the exact 
positioning of the FeGP cofactor and the methenyl-H$_4$MPT$^+$ substrate in the active site. As water molecules are still able to access the active site in 
the {\it closed} conformation (see Sect.\ \ref{sec:QMMM_MD_CLOSED}), we can envisage that the excess proton on the thiol ligand is removed from the active site \textit{via} water.

\section{Discussion and Conclusions}
[FeFe] and [NiFe] hydrogenases cleave or form H$_2$ by redox chemistry \cite{Fontecilla-Camps2007,Vincent2007,Lubitz2007}; a basic 
group close to the active iron atom in [FeFe] hydrogenases is important to donate or accept protons. The mechanism of hydrogen activation 
in [Fe] hydrogenase is different. The enzyme has two large-scale conformations, which differ in the relative orientation
of the central and peripheral subunits. In the {\it closed} conformation, the mononuclear iron cofactor (FeGP) and the substrate are 
kept in close proximity in an arrangement that is stable over longer time scales, as we have shown by MD simulations. Our QM/MM calculations 
have demonstrated that the pyridinol hydroxyl group can easily be deprotonated \textit{via} His14 to form the oxypyridine ligand. The pyridinol 
ligand in [Fe] hydrogenase thus has a function similar to the bridgehead amine group of the H-cluster in [FeFe] hydrogenases \cite{Fan2001}. 
However, the oxypyridine plays an additional, crucial role in activating H$_2$: It is close to the iron atom and represents an ideal Lewis base. 
On the other side of the iron is the carbocationic C14a of the substrate methenyl-H$_4$MPT$^+$, which is an ideal Lewis acid. Furthermore, 
both groups are ionic. When a hydrogen molecule coordinates to the iron, it is polarized by these charges and subjected to an electronic push--pull 
effect exerted by the Lewis pair. 
The spatial arrangement in the {\it closed} conformation is exactly such that the coordinated H$_2$ lies in-between C14a$^+$ and O$^-$.
This leads to facile, exothermic heterolytic H$_2$ cleavage, without involving electron transfers to/from the metal center. The Lewis acid C14a$^+$ is
only present in proximity to FeGP when methenyl-H$_4$MPT$^+$ is bound in the {\it closed} conformation. H$_2$ cleavage in the {\it open} conformation is thus unlikely.

The activation mechanism we have described is reminiscent of hydrogen activation by frustrated Lewis pairs \cite{Stephan2010}. The hydrogen-bound adduct does
not need to be very stable since the H$_2$ cleavage barrier is extremely low (about $1 \, \rm kcal/mol$). Hence, any H$_2$ binding event can 
directly lead to H$_2$ cleavage, without requiring a long-lived H$_2$-bound intermediate.

When the pyridinol ligand is not deprotonated, it is still possible to split H$_2$ {\it via} proton transfer to the thiolate ligand. However, 
we have found this pathway to be much less favorable. This is consistent with observations in biomimetic model complexes that thiol is a poor 
ligand \cite{Chen2012}. The pyridinol/oxypyridine equilibrium must be strongly affected by the pH, so we would expect the reactivity to depend 
critically on pH as well, which is indeed the case \cite{Zirngibl1990,Schwoerer1993}.

The atomistic mechanism of H$_2$ activation in [Fe] hydrogenase we have proposed herein satisfies the criteria set out in Sect.\ \ref{sec:Intro}: 
No stable hydride intermediate; no occurrence of, or requirement for, a long-lived H$_2$ adduct; no involvement of the thiolate ligand 
as a proton acceptor; a crucial role for His14.  In our preferred mechanism, the pyridinol hydroxyl group and His14, together with the stable 
placement of the substrate carbocation in the active site, are the essential players, which is in accord with the observation that the enzyme 
looses $99 \, \%$ of its activity upon H14A 
mutation \cite{Shima2008}. The residual activity of the H14A mutant can be explained by the alternative, less favorable activation pathway {\it via} the thiolate.

In the {\it open} conformation, which might be prevailing in solution, the water-bound FeGP cofactor is the most probable 
form, which agrees with the results of a theoretical M\"ossbauer study \cite{Gubler2013}.

\footnotesize{

\begin{thebibliography}{10}

\bibitem{Schwoerer1993}
Schw{\"o}rer,~B.;\ \ Fernandez,~V.~M.;\ \ Zirngibl,~C.;\ \ Thauer,~R.~K.
  \textit{Eur. J. Biochem.} \textbf{1993,} \textsl{212,} 255--261.

\bibitem{Klein1995}
Klein,~A.~R.;\ \ Hartmann,~G.~C.;\ \ Thauer,~R.~K. \textit{Eur. J. Biochem.}
  \textbf{1995,} \textsl{233,} 372--376.

\bibitem{Hartmann1996}
Hartmann,~G.~C.;\ \ Santamaria,~E.;\ \ Fern{\'a}ndez,~V.~M.;\ \ Thauer,~R.~K.
  \textit{J. Biol. Inorg. Chem.} \textbf{1996,} \textsl{1,} 446--450.

\bibitem{Zirngibl1992}
Zirngibl,~C.;\ \ van Dongen,~W.;\ \ Schw{\"o}rer,~B.;\ \ von B{\"u}nau,~R.;\ \
  Richter,~M.;\ \ Klein,~A.;\ \ Thauer,~R.~K. \textit{Eur. J. Biochem.}
  \textbf{1992,} \textsl{208,} 511--520.

\bibitem{Corr2011}
Corr,~M.~J.;\ \ Murphy,~J.~A. \textit{Chem. Soc. Rev.} \textbf{2011,}
  \textsl{40,} 2279--2292.

\bibitem{Lubitz2014}
Lubitz,~W.;\ \ Ogata,~H.;\ \ R\"udiger,~O.;\ \ Reijerse,~E. \textit{Chem. Rev.}
  \textbf{2014,} \textsl{114,} 4081--4148.

\bibitem{Vincent2007}
Vincent,~K.~A.;\ \ Parker,~A.;\ \ Armstrong,~F.~A. \textit{Chem. Rev.}
  \textbf{2007,} \textsl{107,} 4366--4413.

\bibitem{Fontecilla-Camps2007}
Fontecilla-Camps,~J.~C.;\ \ Volbeda,~A.;\ \ Cavazza,~C.;\ \ Nicolet,~Y.
  \textit{Chem. Rev.} \textbf{2007,} \textsl{107,} 4273--4303.

\bibitem{Lubitz2007}
Lubitz,~W.;\ \ Reijerse,~E.;\ \ van Gastel,~M. \textit{Chem. Rev.}
  \textbf{2007,} \textsl{107,} 4331--4365.

\bibitem{DeLacey2007}
De~Lacey,~A.~L.;\ \ Fern{\'a}ndez,~V.~M.;\ \ Rousset,~M.;\ \ Cammack,~R.
  \textit{Chem. Rev.} \textbf{2007,} \textsl{107,} 4304--4330.

\bibitem{Lyon2004}
Lyon,~E.~J.;\ \ Shima,~S.;\ \ Buurman,~G.;\ \ Chowdhuri,~S.;\ \
  Batschauer,~A.;\ \ Steinbach,~K.;\ \ Thauer,~R.~K. \textit{Eur. J. Biochem.}
  \textbf{2004,} \textsl{271,} 195--204.

\bibitem{Shima2005}
Shima,~S.;\ \ Lyon,~E.~J.;\ \ Thauer,~R.~K.;\ \ Mienert,~B.;\ \ Bill,~E.
  \textit{J. Am. Chem. Soc.} \textbf{2005,} \textsl{127,} 10430--10435.

\bibitem{Wang2008}
Wang,~X.;\ \ Li,~Z.;\ \ Zeng,~X.;\ \ Luo,~Q.;\ \ Evans,~D.~J.;\ \
  Pickett,~C.~J.;\ \ Liu,~X. \textit{Chem. Commun.} \textbf{2008,} \textsl{30,}
  3555--3557.

\bibitem{Salomone-Stagni2010}
Salomone-Stagni,~M.;\ \ Stellato,~F.;\ \ Whaley,~C.~M.;\ \ Vogt,~S.;\ \
  Morante,~S.;\ \ Shima,~S.;\ \ Rauchfuss,~T.~B.;\ \ Meyer-Klaucke,~W.
  \textit{Dalton Trans.} \textbf{2010,} \textsl{39,} 3057--3064.

\bibitem{Lyon2004a}
Lyon,~E.~J.;\ \ Shima,~S.;\ \ Boecher,~R.;\ \ Thauer,~R.~K.;\ \
  Grevels,~F.-W.;\ \ Bill,~E.;\ \ Roseboom,~W.;\ \ Albracht,~S.~P. \textit{J.
  Am. Chem. Soc.} \textbf{2004,} \textsl{126,} 14239--14248.

\bibitem{Hiromoto2009a}
Hiromoto,~T.;\ \ Warkentin,~E.;\ \ Moll,~J.;\ \ Ermler,~U.;\ \ Shima,~S.
  \textit{Angew. Chem. Int. Ed.} \textbf{2009,} \textsl{48,} 6457--6460.

\bibitem{Yang2009}
Yang,~X.;\ \ Hall,~M.~B. \textit{J. Am. Chem. Soc.} \textbf{2009,}
  \textsl{131,} 10901--10908.

\bibitem{Finkelmann2013a}
Finkelmann,~A.~R.;\ \ Stiebritz,~M.~T.;\ \ Reiher,~M. \textit{J. Phys. Chem. B}
  \textbf{2013,} \textsl{117,} 4806--4817.

\bibitem{Gubler2013}
Gubler,~J.;\ \ Finkelmann,~A.~R.;\ \ Reiher,~M. \textit{Inorg. Chem.}
  \textbf{2013,} \textsl{52,} 14205-14215.

\bibitem{Hedegard2014}
Hedeg{\aa}rd,~E.~D.;\ \ Knecht,~S.;\ \ Ryde,~U.;\ \ Kongsted,~J.;\ \ Saue,~T.
  \textit{Phys. Chem. Chem. Phys.} \textbf{2014,} .

\bibitem{Chen2011}
Chen,~D.;\ \ Scopelliti,~R.;\ \ Hu,~X. \textit{Angew. Chem. Int. Ed.}
  \textbf{2011,} \textsl{50,} 5671--5673.

\bibitem{Chen2011a}
Chen,~D.;\ \ Ahrens-Botzong,~A.;\ \ Sch\"unemann,~V.;\ \ Scopelliti,~R.;\ \
  Hu,~X. \textit{Inorg. Chem.} \textbf{2011,} \textsl{50,} 5249--5257.

\bibitem{Chen2010}
Chen,~D.;\ \ Scopelliti,~R.;\ \ Hu,~X. \textit{Angew. Chem. Int. Ed.}
  \textbf{2010,} \textsl{49,} 7512--7515.

\bibitem{Chen2012}
Chen,~D.;\ \ Scopelliti,~R.;\ \ Hu,~X. \textit{Angew. Chem.} \textbf{2012,}
  \textsl{51,} 1955--1957.

\bibitem{Shima2008}
Shima,~S.;\ \ Pilak,~O.;\ \ Vogt,~S.;\ \ Schick,~M.;\ \ Stagni,~M.~S.;\ \
  Meyer-Klaucke,~W.;\ \ Warkentin,~E.;\ \ Thauer,~R.~K.;\ \ Ermler,~U.
  \textit{Science} \textbf{2008,} \textsl{321,} 572--575.

\bibitem{Pilak2006}
Pilak,~O.;\ \ Mamat,~B.;\ \ Vogt,~S.;\ \ Hagemeier,~C.~H.;\ \ Thauer,~R.~K.;\ \
  Shima,~S.;\ \ Vonrhein,~C.;\ \ Warkentin,~E.;\ \ Ermler,~U. \textit{J. Mol.
  Biol.} \textbf{2006,} \textsl{358,} 798--809.

\bibitem{Hiromoto2009}
Hiromoto,~T.;\ \ Ataka,~K.;\ \ Pilak,~O.;\ \ Vogt,~S.;\ \ Stagni,~M.~S.;\ \
  Meyer-Klaucke,~W.;\ \ Warkentin,~E.;\ \ Thauer,~R.~K.;\ \ Shima,~S.;\ \
  Ermler,~U. \textit{FEBS Lett.} \textbf{2009,} \textsl{583,} 585--590.

\bibitem{Warshel1976}
Warshel,~A.;\ \ Levitt,~M. \textit{J. Mol. Biol.} \textbf{1976,} \textsl{103,}
  227--249.

\bibitem{Senn2009}
Senn,~H.~M.;\ \ Thiel,~W. \textit{Angew. Chem. Int. Ed.} \textbf{2009,}
  \textsl{48,} 1198--1229.

\bibitem{Wang2004}
Wang,~J.;\ \ Wolf,~R.~M.;\ \ Caldwell,~J.~W.;\ \ Kollman,~P.~A.;\ \ Case,~D.~A.
  \textit{J. Comput. Chem.} \textbf{2004,} \textsl{25,} 1157--1174.

\bibitem{Tolman1938}
Tolman,~R.~C. \textit{The principles of statistical mechanics;} Oxford
  University Press: London, UK: 1938.

\bibitem{Duan2003}
Duan,~Y.;\ \ Wu,~C.;\ \ Chowdhury,~S.;\ \ Lee,~M.~C.;\ \ Xiong,~G.;\ \
  Zhang,~W.;\ \ Yang,~R.;\ \ Cieplak,~P.;\ \ Luo,~R.;\ \ Lee,~T.;\ \
  Caldwell,~J.;\ \ Wang,~J.;\ \ Kollman,~P. \textit{J. Comput. Chem.}
  \textbf{2003,} \textsl{24,} 1999--2012.

\bibitem{Lee2004}
Lee,~M.~C.;\ \ Duan,~Y. \textit{Proteins} \textbf{2004,} \textsl{55,} 620--634.

\bibitem{Berendsen1995}
Berendsen,~H. J.~C.;\ \ van~der Spoel,~D.;\ \ van Drunen,~R. \textit{Comput.
  Phys. Commun.} \textbf{1995,} \textsl{91,} 43--56.

\bibitem{vanderSpoel2005}
Van Der~Spoel,~D.;\ \ Lindahl,~E.;\ \ Hess,~B.;\ \ Groenhof,~G.;\ \
  Mark,~A.~E.;\ \ Berendsen,~H. J.~C. \textit{J. Comput. Chem.} \textbf{2005,}
  \textsl{26,} 1701--1718.

\bibitem{Hess2008}
Hess,~B.;\ \ Kutzner,~C.;\ \ van~der Spoel,~D.;\ \ Lindahl,~E. \textit{J. Chem.
  Theor. Comput.} \textbf{2008,} \textsl{4,} 435--447.

\bibitem{Pronk2013}
Pronk,~S.;\ \ Páll,~S.;\ \ Schulz,~R.;\ \ Larsson,~P.;\ \ Bjelkmar,~P.;\ \
  Apostolov,~R.;\ \ Shirts,~M.~R.;\ \ Smith,~J.~C.;\ \ Kasson,~P.~M.;\ \
  van~der Spoel,~D.;\ \ Hess,~B.;\ \ Lindahl,~E. \textit{Bioinformatics}
  \textbf{2013,} \textsl{29,} 845-854.

\bibitem{chemshell}
``ChemShell, a Computational Chemistry Shell'',  see www.chemshell.org.

\bibitem{Sherwood2003}
Sherwood,~P. \textit{et al.}\  \textit{Theochem} \textbf{2003,} \textsl{632,}
  1--28.

\bibitem{Metz2014}
Metz,~S.;\ \ K\"astner,~J.;\ \ Sokol,~A.~A.;\ \ Keal,~T.~W.;\ \ Sherwood,~P.
  \textit{WIRS Comput. Mol. Sci.} \textbf{2014,} \textsl{4,} 101--110.

\bibitem{Ahlrichs1989}
Ahlrichs,~R.;\ \ B{\"a}r,~M.;\ \ H{\"a}ser,~M.;\ \ Horn,~H.;\ \ K{\"o}lmel,~C.
  \textit{Chem. Phys. Lett.} \textbf{1989,} \textsl{162,} 165--169.

\bibitem{Turbomole65}
``{TURBOMOLE V6.5 2013}, a development of {University of Karlsruhe} and
  {Forschungszentrum Karlsruhe GmbH}, 1989-2007, {TURBOMOLE GmbH}, since 2007;
  available from \\ {\tt http://www.turbomole.com}.'', .

\bibitem{Tao2003}
Tao,~J.;\ \ Perdew,~J.~P.;\ \ Staroverov,~V.~N.;\ \ Scuseria,~G.~E.
  \textit{Phys. Rev. Lett.} \textbf{2003,} \textsl{91,} 146401.

\bibitem{Grimme2010}
Grimme,~S.;\ \ Antony,~J.;\ \ Ehrlich,~S.;\ \ Krieg,~H. \textit{J. Chem. Phys.}
  \textbf{2010,} \textsl{132,} 154104.

\bibitem{Weigend2005}
Weigend,~F.;\ \ Ahlrichs,~R. \textit{Phys. Chem. Chem. Phys.} \textbf{2005,}
  \textsl{7,} 3297--3305.

\bibitem{Schaefer1992}
Sch{\"a}fer,~A.;\ \ Horn,~H.;\ \ Ahlrichs,~R. \textit{J. Chem. Phys.}
  \textbf{1992,} \textsl{97,} 2571.

\bibitem{Fan2001}
Fan,~H.-J.;\ \ Hall,~M.~B. \textit{J. Am. Chem. Soc.} \textbf{2001,}
  \textsl{123,} 3828--3829.

\bibitem{Stephan2010}
Stephan,~D.~W.;\ \ Erker,~G. \textit{Angew. Chem. Int. Ed.} \textbf{2010,}
  \textsl{49,} 46--76.

\bibitem{Zirngibl1990}
Zirngibl,~C.;\ \ Hedderich,~R.;\ \ Thauer,~R.~K. \textit{FEBS Lett.}
  \textbf{1990,} \textsl{261,} 112--116.

\end{thebibliography}
\providecommand{\refin}[1]{\\ \textbf{Referenced in:} #1}

}




\section*{Supplementary material}

# Supporting Information

# Hydrogen-Activation Mechanism of [Fe] Hydrogenase Revealed by Multi-Scale Modeling

Arndt Robert Finkelmann[a], Hans Martin Senn[b]*, Markus Reiher[a]†

[a] ETH Zürich, Laboratorium für Physikalische Chemie, Vladimir-Prelog-Weg 2
8093 Zürich, Switzerland

[b] WestCHEM and School of Chemistry, University of Glasgow, Glasgow G12 8QQ, UK

May 31, 2014

*Corresponding author; e-mail: hans.senn@glasgow.ac.uk
†Corresponding author; e-mail: markus.reiher@phys.chem.ethz.ch

# 1 Computational details

## 1.1 Parametrization of methylene-tetrahydromethanopterin and iron-guanylylpyridinol

To be able to study [Fe] hydrogenase by means of molecular-dynamics (MD) simulations and quantum-mechanics/molecular-mechanics (QM/MM) calculations a force-field description of the whole system is necessary. We chose the Amber force field (version ff03 [1,2]) for the molecular-mechanics (MM) part. This force field can be combined with the general Amber force field (GAFF),[3] which allows for an easy parametrization of organic molecules. We followed the GAFF parametrization procedure to derive parameters for the iron-guanylylpyridinol (FeGP) (see Fig. 1 in the main paper) cofactor and the substrate methylene-tetrahydromethanopterin (methylene-$H_4$MPT, see Fig. 1 in the main paper). The parametrization can be divided into two steps. As first step, partial charges derived from the electrostatic potential (ESP) for all atoms were calculated. Then, GAFF topology files for the organic parts were created. The parameters used are given in Figs. 1, 2 and Tables 1, 2.

### 1.1.1 Calculation of partial charges

For the calculation of partial charges, we chose a procedure similar to that applied in Ref. 1. Structures were optimized with the GAUSSIAN09 program package,[4] utilizing density functional theory with the B3LYP exchange-correlation functional,[5–7] a 6-31+G* basis set[8,9] and the integral equation formalism polarizable continuum model (IEFPCM)[10,11] with $\epsilon = 4$ to account for electrostatic screening. Note that in Ref. 1 structures were optimized with the Hartree–Fock method. Partial charges were calculated with the TURBOMOLE program package (version 6.3.1)[12,13] according to a scheme related to that suggested by Kollman.[14] For the TURBOMOLE calculations, density functional theory with the B3LYP exchange-correlation functional,[5–7] the conductor-like screening model ($\epsilon = 4$)[15] and an aug-cc-pVTZ basis set[16–18] on all atoms was employed.

Unconstrained structure optimization of methenyl-$H_4$MPT (starting structure from PDB file 3H65[19]) leads to the formation of internal hydrogen bonds of the tailing carboxylate groups to hydroxo groups of either the furanose ring or the glycerin-derived part. These internal hydrogen bonds are not found in the protein structure because the carboxylates can form intermolecular hydrogen bonds. Hence, intramolecular hydrogen bonds do not resemble the bonding situation in the crystal structure and need to be avoided. This was achieved by constraining several dihedral angles during optimization.

### 1.1.2 Generation of GAFF topologies

The generation of GAFF topologies for methylene-$H_4$MPT was straightforward. Topology files were created with the program ACPYPE,[20] which interfaces ANTECHAMBER[21] of AMBERTOOLS 13[22] to create AMBER and GROMACS topology files. The whole methylene-$H_4$MPT molecule was treated as one MM-residue.

Generation of suitable parameters for the FeGP cofactor was more involved.[23] Since metal complexes are not parametrizable with GAFF the cofactor was split up into the guanylylpyridinol ligand, 2 CO molecules, the iron ion (each treated

as one MM-residue) and the cysteine rest. For the iron ion a new ion type was introduced. The two CO molecules were parametrized with GAFF. The guanylylpyridinol ligand could be parametrized with GAFF, however, the parameters of the acyl part coordinating the iron atom had to be adjusted (equilibrium value of the acyl O–C–C angle) (see Fig. 1 in the main paper for the structure). In the crystal structure, the open coordination site is coordinated by water. For this water molecule, the standard parameters of the TIP3P water model[24] were used as the ESP-calculated partial charges differed insignificantly from the standard charges. For the cysteinate rest, a modified cysteinate residue was defined, which had the same bonded and van der Waals parameters as a standard cysteinate, but ESP-derived charges. With a view on terminating the QM region at the cysteinate $C^\beta$–$C^\alpha$ bond in the subsequent QM/MM calculations, while maintaining integer charges for the QM and MM regions, a charge of $+0.069588e$ was distributed equally among the $C^\beta H_2$ atoms of the new cysteinate residue to obtain a total integer charge of $-1e$ for FeGP. Since the Fe atom, both CO molecules, the cysteine sulfur atom, and the CO atoms of the acyl-ligand were positionally restrained during the classical MD simulations, no metal–ligand bonded parameters had to be derived.

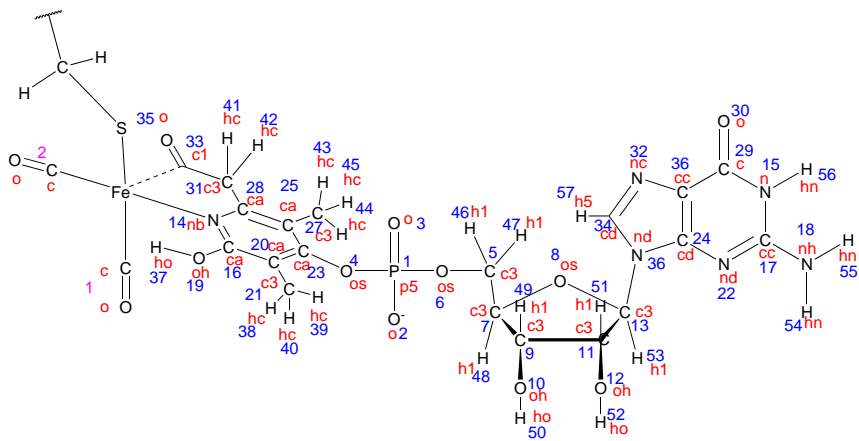

Figure 1: Atom types and atom numbers for the FeGP cofactor. GAFF atom types are given in red, atom numbers in blue. Pink numbers indicate the first and the second CO ligand.

## 1.2 Preparation of the protein structures for MD simulations

### 1.2.1 *Open* conformation

The simulations are based on the crystal structure published by Hiromoto *et al.*[19] (PDB code: 3H65). They achieved to crystallize a C176A mutant that harbors the cofactor, ligated by dithiothreitol (buffer molecule) instead of Cys176, in complex with the substrate methylene-$H_4$MPT. The crystal structure is in the *open* conformation. In the crystal structure, the C-terminal residues 346–358

3

Table 1: Atom types and partial charges of the FeGP cofactor for the MD simulations and for the QM/MM calculations. The structure with the corresponding atom numbers is given in Fig. 1.

| atom | MD charge | QMMM charge || atom | MD charge | QMMM charge |
|---|---|---|---|---|---|
| **Fe(CO)$_2$SC$^\beta$H$_2$** | | | | | |
| Fe | 0.7416375 | 0.741585 || S | −0.4603 | −0.460350 |
| C1 | 0.074899 | 0.074849 || C$^\beta$ | −0.186503 | −0.186553 |
| O1 | −0.223267 | −0.223317 || H$^\beta$ | 0.126296 | 0.126246 |
| C2 | 0.143396 | 0.143346 || H$^\beta$ | 0.126296 | 0.126246 |
| O2 | −0.243743 | −0.243793 || | | |
| **guanylylpyridinol** | | | | | |
| 1 p5 | 1.173428 | 1.173379 || 30 o | −0.638592 | −0.638525 |
| 2 o | −0.804905 | −0.804954 || 31 c3 | −0.290532 | −0.290582 |
| 3 o | −0.758151 | −0.758200 || 32 nc | −0.659167 | −0.659100 |
| 4 os | −0.248012 | −0.248061 || 33 c1 | 0.510845 | 0.510795 |
| 5 c3 | −0.232453 | −0.232502 || 34 cd | 0.214577 | 0.214644 |
| 6 os | −0.451749 | −0.451798 || 35 o | −0.634698 | −0.634748 |
| 7 c3 | 0.336367 | 0.336435 || 36 na | −0.102300 | −0.102233 |
| 8 os | −0.579760 | −0.579692 || 37 ho | 0.456286 | 0.456236 |
| 9 c3 | −0.019723 | −0.019655 || 38 hc | 0.127078 | 0.127028 |
| 10 oh | −0.646053 | −0.645985 || 39 hc | 0.164999 | 0.164949 |
| 11 c3 | 0.161410 | 0.161478 || 40 hc | 0.168804 | 0.168754 |
| 12 oh | −0.687887 | −0.687819 || 41 hc | 0.148313 | 0.148263 |
| 13 c3 | 0.430294 | 0.430362 || 42 hc | 0.124692 | 0.124642 |
| 14 nb | −0.296088 | −0.296137 || 43 hc | 0.100120 | 0.100070 |
| 15 n | −0.657113 | −0.657045 || 44 hc | 0.147854 | 0.147804 |
| 16 ca | 0.364888 | 0.364839 || 45 hc | 0.130894 | 0.130844 |
| 17 cc | 0.770287 | 0.770355 || 46 h1 | 0.120443 | 0.120393 |
| 18 nh | −0.851084 | −0.851016 || 47 h1 | 0.172266 | 0.172216 |
| 19 oh | −0.601450 | −0.601499 || 48 h1 | 0.088891 | 0.088958 |
| 20 ca | 0.244988 | 0.244939 || 49 h1 | 0.235349 | 0.235416 |
| 21 c3 | −0.501514 | −0.501563 || 50 ho | 0.408312 | 0.408379 |
| 22 nd | −0.690035 | −0.689967 || 51 h1 | 0.093079 | 0.093146 |
| 23 ca | −0.222549 | −0.222598 || 52 ho | 0.439216 | 0.439283 |
| 24 cd | 0.291468 | 0.291536 || 53 h2 | 0.035636 | 0.035703 |
| 25 ca | 0.385152 | 0.385101 || 54 hn | 0.400592 | 0.400659 |
| 26 cc | 0.184629 | 0.184696 || 55 hn | 0.389254 | 0.389321 |
| 27 c3 | −0.432668 | −0.432718 || 56 hn | 0.415823 | 0.415890 |
| 28 ca | −0.163103 | −0.163153 || 57 h5 | 0.138101 | 0.138168 |
| 29 c | 0.595243 | 0.595310 || | | |
| **H$_2$O** | | | | | |
| O | −0.834000 | −0.834000 || H | 0.417000 | 0.417000 |
| H | 0.417000 | 0.417000 || | | |

Table 2: Atom types and partial charges of methylene-H$_4$MPT for the MD simulations and for the QM/MM calculations. The structure with the corresponding atom numbers is given in Fig. 2.

| atom  | MD charge  | QMMM charge  | atom  | MD charge  | QMMM charge  |
|-------|------------|--------------|-------|------------|--------------|
| 1 nc  | −0.839624  | −0.839569    | 49 o  | −0.918099  | −0.918141    |
| 2 cd  | 0.866549   | 0.866604     | 50 c3 | −0.462739  | −0.462781    |
| 3 nh  | −0.869435  | −0.869380    | 51 c3 | 0.054124   | 0.054082     |
| 4 n   | −0.711463  | −0.711408    | 52 c  | 0.877807   | 0.877765     |
| 5 c   | 0.716700   | 0.716755     | 53 o  | −0.876948  | −0.876990    |
| 6 o   | −0.679835  | −0.679780    | 54 o  | −0.944601  | −0.944643    |
| 7 cd  | −0.258910  | −0.258855    | 55 h1 | 0.107622   | 0.107580     |
| 8 nh  | −0.378333  | −0.378278    | 56 h1 | 0.049829   | 0.049787     |
| 9 c3  | 0.049297   | 0.049352     | 57 hc | 0.025634   | 0.025592     |
| 10 c3 | 0.358361   | 0.358415     | 58 hc | 0.086049   | 0.086007     |
| 11 c3 | −0.632035  | −0.631981    | 59 hc | −0.001997  | −0.002039    |
| 12 nh | −0.584658  | −0.584604    | 60 hc | 0.017822   | 0.017780     |
| 13 cc | 0.633917   | 0.633971     | 61 hc | 0.121244   | 0.121202     |
| 14 c3 | 0.474230   | 0.474284     | 62 hc | 0.094270   | 0.094228     |
| 15 c3 | −0.473377  | −0.473323    | 63 ho | 0.417555   | 0.417513     |
| 16 c3 | −0.006172  | −0.006118    | 64 ho | 0.402850   | 0.402808     |
| 17 nh | −0.348532  | −0.348478    | 65 ho | 0.364890   | 0.364848     |
| 18 ca | 0.224733   | 0.224787     | 66 ho | 0.434832   | 0.434790     |
| 19 ca | −0.327418  | −0.327364    | 67 ho | 0.440181   | 0.440139     |
| 20 ca | −0.277733  | −0.277679    | 68 hn | 0.384391   | 0.384445     |
| 21 ca | 0.305265   | 0.305319     | 69 hn | 0.392875   | 0.392929     |
| 22 ca | −0.231917  | −0.231863    | 70 hc | 0.088181   | 0.088235     |
| 23 ca | −0.320717  | −0.320663    | 71 hc | 0.118277   | 0.118331     |
| 24 c3 | −0.214567  | −0.214610    | 72 hc | 0.129347   | 0.129401     |
| 25 c3 | 0.490123   | 0.490080     | 73 hc | 0.152460   | 0.152514     |
| 26 c3 | 0.131906   | 0.131863     | 74 hc | 0.165743   | 0.165797     |
| 27 c3 | 0.268650   | 0.268607     | 75 hc | 0.170784   | 0.170838     |
| 28 c3 | 0.007070   | 0.007027     | 76 h1 | 0.075703   | 0.075661     |
| 29 oh | −0.639264  | −0.639307    | 77 h1 | 0.107458   | 0.107416     |
| 30 oh | −0.670502  | −0.670545    | 78 h2 | 0.164069   | 0.164123     |
| 31 oh | −0.677646  | −0.677689    | 79 h2 | 0.045246   | 0.045300     |
| 32 os | −0.562346  | −0.562389    | 80 h1 | 0.070626   | 0.070584     |
| 33 os | −0.440433  | −0.440475    | 81 h1 | 0.013312   | 0.013270     |
| 34 c3 | −0.048424  | −0.048466    | 82 h1 | −0.018889  | −0.018931    |
| 35 c3 | 0.161302   | 0.161260     | 83 h1 | 0.008358   | 0.008316     |
| 36 os | −0.512516  | −0.512558    | 84 hn | 0.348778   | 0.348832     |
| 37 c3 | 0.653326   | 0.653284     | 85 hn | 0.414010   | 0.414064     |
| 38 oh | −0.750097  | −0.750139    | 86 h1 | −0.062031  | −0.061977    |
| 39 c3 | −0.038881  | −0.038923    | 87 h1 | 0.039663   | 0.039717     |
| 40 oh | −0.748423  | −0.748465    | 88 h1 | 0.081657   | 0.081711     |
| 41 c3 | 0.577317   | 0.577275     | 89 h1 | 0.018267   | 0.018225     |
| 42 p5 | 1.257785   | 1.257743     | 90 h1 | −0.094589  | −0.094631    |
| 43 o  | −0.893733  | −0.893775    | 91 h1 | 0.095968   | 0.095926     |
| 44 o  | −0.855874  | −0.855916    | 92 h2 | 0.054208   | 0.054166     |
| 45 os | −0.423072  | −0.423114    | 93 ha | 0.180798   | 0.180852     |
| 46 c3 | 0.299822   | 0.299780     | 94 ha | 0.146830   | 0.146884     |
| 47 c  | 0.923333   | 0.923291     | 95 ha | 0.165934   | 0.165988     |
| 48 o  | −0.913326  | −0.913368    | 96 ha | 0.181818   | 0.181872     |

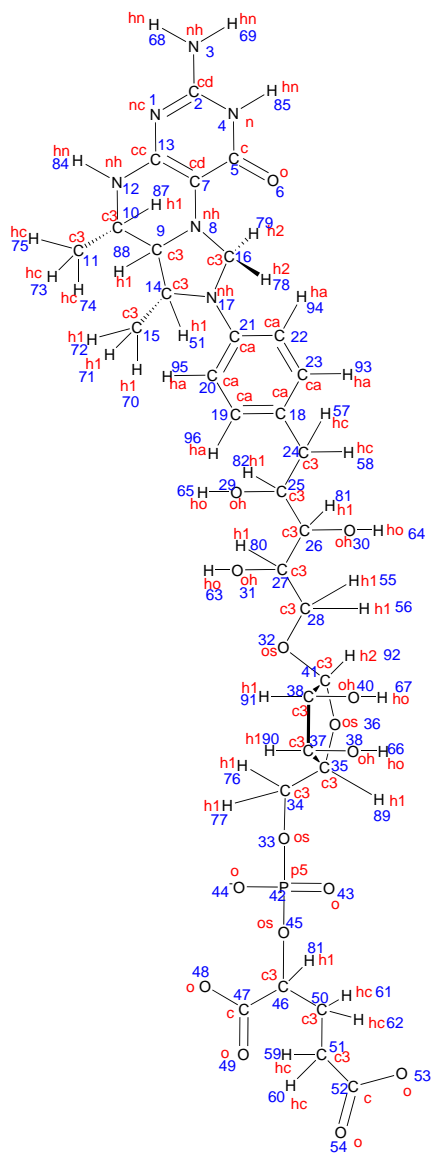

Figure 2: Atom types and atom numbers of methylene-H$_4$MPT. GAFF atom types are given in red, atom numbers in blue.

could not be refined. They form an unordered tail, far away from the reactive centers, and thus were simply omitted. To prepare the structure for the simulations, several modifications had to be made. FeGP is ligated by dithiothreitol in the crystal structure, which leads to a slight displacement of the whole cofactor compared to its position in the wild-type structure (without methylene-H$_4$MPT, PDB code: 3F47[25]). To obtain a wild-type-like structure Ala176 was mutated

Table 3: Partial charges of His14 for the QM/MM calculations. Atom names are according to the Amber ff03.

| atom | QM/MM charge |
|------|-------------:|
| N    | −0.509660    |
| H    | 0.348160     |
| CA   | 0.116205     |
| HA   | 0.134900     |
| CB   | −0.121078    |
| HB2  | 0.087889     |
| HB3  | 0.087889     |
| CG   | 0.000013     |
| ND1  | −0.204225    |
| HD1  | 0.319833     |
| CE1  | 0.148833     |
| HE1  | 0.123742     |
| NE2  | −0.599922    |
| CD2  | 0.045304     |
| HD2  | 0.111717     |
| C    | 0.513086     |
| O    | −0.602692    |

Table 4: Protonation states of His residues of the modified 3H65 crystal structure for MD simulations, as determined with the program REDUCE.[27] $\delta$ indicates a proton at $N^\delta$ and $\epsilon$ indicates a proton at $N^\epsilon$. pKa values were calculated with PROPKA.[28]

| residue | H position | pKa value |
|---------|:----------:|----------:|
| HIS 14  | $\delta$   | 6.57      |
| HIS 41  | $\delta$   | 0.60      |
| HIS 52  | $\delta$   | 5.68      |
| HIS 88  | $\delta$   | 5.24      |
| HIS 120 | $\epsilon$ | 3.99      |
| HIS 123 | $\epsilon$ | 5.82      |
| HIS 174 | $\delta$   | 4.08      |
| HIS 201 | $\epsilon$ | 1.68      |
| HIS 294 | $\epsilon$ | 5.82      |
| HIS 340 | $\delta$   | 7.22      |

back to cysteinate and the dithiothreitol ligand was removed from the structure. To place the cofactor in the correct (wild-type) position, the structure of the wild type (3F47) was aligned with the structure of the C176A mutant (3H65) using PyMOL.[26] The coordinates of the FeGP moiety (including the cysteine S atom) in the aligned wild-type structure were then inserted into the modified mutant structure, replacing those of the misplaced FeGP. With this procedure, the position of FeGP in the modified mutant structure resembled the position in the wild-type structure. Only the $C^\beta$–S bond of Cys176 (formerly Ala176) was stretched from 1.79 Å in the wild-type structure to 2.21 Å in the modified mutant structure. The correct bond length is restored in the energy-minimization step at the start of the MD simulations. The protonation states of titratable residues were determined with the program REDUCE,[27] which can also correct flipped Asn/Gln/His residues (none were in this case), and verified with PROPKA.[28] All titratable residues were in their standard protonation states; His protonation states are summarized in Table 4. Finally, the whole

protein dimer was created from the monomer chain with PyMOL.[26]

### 1.2.2 *Closed* conformation

The *closed* conformation was generated from the modified crystal structure of the *open* conformation (see previous section), which we call starting structure here. The only crystal structure available in the *closed* conformation is for the wild-type apoenzyme (PDB code: 2B0J[29]). To build a model of the complete enzyme in the *closed* conformation, the central subunit (residues 253–345) of 2B0J was first aligned to the central subunit of the starting structure. In the second step, the peripheral subunit (residues 1–241) of the starting structure was aligned to the peripheral subunit of the now aligned 2B0J. The FeGP cofactor was aligned together with the peripheral subunit. The structure thus obtained for the *closed* conformation of the holoenzyme–substrate complex had no significant atom overlaps, except for the tail part of methylene-$H_4$MPT (after the ribitol part, see Fig. 1 of the main paper). This tail was rotated with the help of the UCSF Chimera program[30] to remove atom overlaps. Given the high flexibility and mobility of the tail, any memory of the initial conformation will be lost during the MD sampling. In the structure thus generated, there is a gap in the backbone between residues 241 and 242, between the hinge region (residues 242–252) and peripheral subunit (see Fig. 3), which will be closed during energy minimization. Finally, the water molecule coordinated to Fe was removed because it would prevent the hydride transfer reaction that we intend to study. With the water molecule absent, the structure is exactly the product of the hydride transfer. The full dimer was again created with PyMOL.[26] A superposition of the starting structure, the final structure, and the apoenzyme in the *open* conformation is presented in Fig. 3. In the generated structure of the *closed* conformation, the distance between Fe and the hydride accepting carbon atom (C14a) is 3.23 Å, which compares well to the distance of 3 Å found by Hiromoto *et al.*, who modelled the structure of the *closed* conformation in a similar fashion.[19]

### 1.3 MD simulation protocol

All MD preparation steps and simulations were performed with the Gromacs molecular dynamics package version 4.5.5.[31–34] The protein was centered in a triclinic box with a minimal distance of 1 nm between solute and box border. The box was solvated and ions ($Na^+$ and $Cl^-$) were added to neutralize the system and to obtain an ion concentration of 0.15 mmol/L. The Fe atom, both CO ligands, the cysteinate S atom, and the CO group of the Fe-coordinating acyl ligand in the FeGP cofactor were kept frozen or positionally restrained at their positions in the prepared crystal structure. Positional restraints, rather than constraints were necessary for the pressure equilibration because pressure scaling with constrained (frozen) atoms leads to technical difficulties in the Gromacs implementation.

The integration time step was 2 fs. The linear constraint solver (LINCS) algorithm to 4th order with 1 iteration was invoked to enforce constraints (all bonds after energy minimization). Water molecules were kept rigid with the SETTLE algorithm in all steps after energy minimization. For neighbor searching, a grid-based group cut-off scheme was used with a cut-off distance of 1 nm

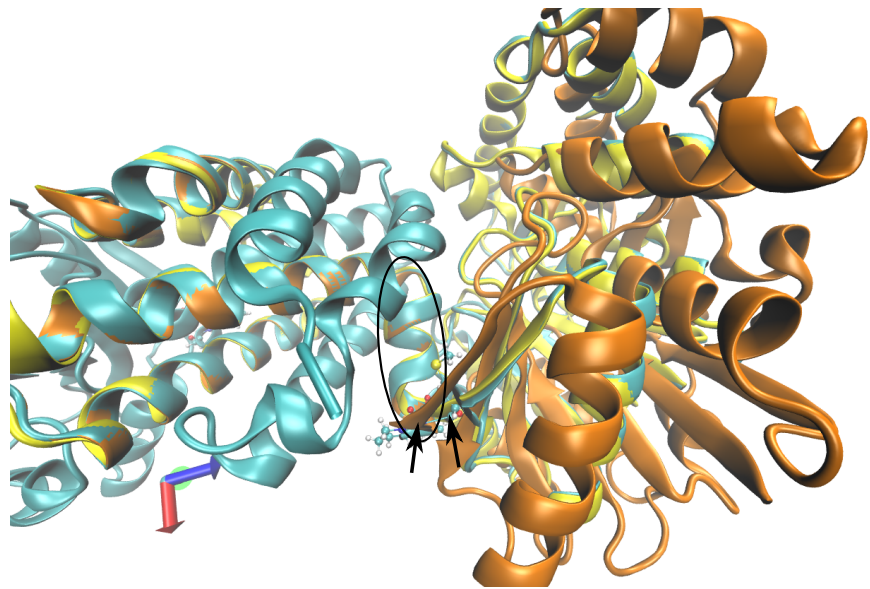

Figure 3: Superposition of the generated *closed* conformation (cyan), the apoenzyme in *closed* conformation (PDB code: 2B0J, yellow) and holoenzyme-substrate complex in *open conformation* (PDB code: 3H65, orange). The central (left side) units are aligned. The peripheral unit of the generated *closed* conformation cyan overlaps with the peripheral unit of the apoenzyme in *closed* conformation. The hinge region is marked by a black oval. The black arrows mark the gap between backbone atoms of residues 241 and 242 (see text).

for short-range interactions and the neighbor list was updated every 10 steps. Coulomb interactions were calculated with a smooth particle–mesh Ewald algorithm with interpolation order of 4, a Coulomb cut-off of 1 nm, Fourier spacing of 0.12 nm, tolerance of $10^{-5}$ and optimized Fourier transforms. Van der Waals interactions were calculated with a cut-off scheme (radius = 1 nm). Energy and pressure were corrected for long-range dispersion effects. Initial velocities were generated according to a Maxwell–Boltzmann distribution at 293 K. For temperature scaling the system was coupled to a $v$-rescale thermostat.[35] During equilibration, several subsystems were coupled to their own thermostats. For production, the entire system was coupled to one thermostat with a relaxation time of 2 ps and reference temperature of 293 K. Box equilibration was achieved with isotropic box rescaling by coupling the system to a Berendsen barostat with 1 ps relaxation time, compressibility of $4.5 \cdot 10^{-5}\,\mathrm{bar}^{-1}$, center-of-mass scaling of reference coordinates, and a target pressure of 1 bar.

The simulation stages are summarized in Table 5. The system in the *open* conformation was energy minimized in vacuum and in solvent. During 200 ps heating in an $NVT$ ensemble, the protein, the cofactor, the substrate and the solvent were coupled to a temperature bath while all heavy atoms of the protein were positionally restrained. Thereafter, the box was equilibrated in an $NPT$ simulation where the restraints on protein atoms were reduced in three

9

steps (total 900 ps). In the first step, protein, cofactor, substrate and solvent were coupled to three thermostats. In the second and third steps, only two thermostats were utilized (solvent, rest of the system). After adjustment of the box vectors (see Table 6), the system was equilibrated in an $NVT$ ensemble for 400 ps with two thermostats (solvent, rest of the system) and finally 2 ns with one thermostat. The production trajectory was 100 ns. Coordinates were saved to disk every 20 ps.

The simulation procedure for the *closed* conformation was similar. Differences are a reduced force threshold for the vacuum minimization, position restraints for the heavy protein atoms during solvent minimization and only two thermostats (solvent, rest of the system) in all box-equilibration simulations. The equilibrated box parameters are collected in Table 6. The final equilibration simulation ($NVT$) had to be elongated for an additional 2 ns with a reduced temperature coupling constant and reduced restraints for Fe and its first-shell ligands (see Table 6) to avoid instabilities in the production run. Still, the production run terminated at 94.6 ns. We trace this back to the initial strain put into the system during crystal structure manipulation, which could not fully relax because the Fe atoms of the two FeGP cofactors were positionally restrained. Furthermore, the parameters for FeGP and methylene-$H_4$MPT might be not optimal for long simulations. However, the trajectory of 94.6 ns is already longer than trajectories normally produced to prepare QM/MM calculations[36] and is sufficiently long to allow us to analyse visited conformations, structural behavior, and protein dynamics around the active site. Trajectories were analysed with the VMD program.[37]

## 1.4 QM/MM setup

For the QM/MM calculations, several snasphots representing important conformations were selected, as described in the main paper. As QM/MM calculations under periodic boundary conditions are not supported by CHEMSHELL, water molecules and ions outside a shell of 18 Å around the quantum-mechanics (QM) region were therefore discarded to create a finite system. The entire protein dimer was retained. QM/MM optimizations were carried out with the CHEMSHELL program[38–40] (version 3.5.0) with TURBOMOLE (version 6.5)[12,41] as the QM back-end. The optimizations were performed in hybrid delocalized internal coordinates (HDLCs) using the HDLCOpt module.[42] The scaling factor for Cartesian coordinates when constructing HDLCs was set to 0.8; the interval to update the pair list and regenerate HDLCs was set to 100 steps; the convergence threshold was set to 0.001 a.u.

Several regions were defined for the QM/MM optimizations. The *QM region* contained all quantum-mechanically described atoms (52 to 84 atoms). The *MM region* contained all other atoms of the system (approximately 12350 to 12900 atoms). Only the atoms in the *active region* (around 670 to 1700 atoms) were allowed to move in structure optimizations. We used a QM/MM microiterative optimization scheme 43, in which the *inner region* (around 61 to 93 atoms) contained the QM atoms, the MM boundary atoms, and the MM atoms bonded to the them.

The QM part of the calculations was treated with the TPSS exchange—correlation functional[44–46] plus Grimme's DFT-D3 dispersion correction.[47] Structures were optimized with the def2-TZVP basis set[48] on iron and the def2-SVP

10

Table 5: Simulation stages for simulation of [Fe] hydrogenase in the *open* conformation (top part) and in the *closed* conformation (bottom part). See text for more details.

| | Minimization vac. | Minimization solv. | Heating | $p$ equilibration | $T$ equilibration | Production |
|---|---|---|---|---|---|---|
| No. of steps | $F_{\text{tol}} = 2000$ kJmol$^{-1}$nm$^{-1}$ | $F_{\text{tol}} = 1000$ kJmol$^{-1}$nm$^{-1}$ | 100000 | 200000 + 50000 +200000 | 200000 +1000000 | 50000000 |
| Propagation | Steepest descent | Steepest descent | | Leap-frog Verlet | | |
| Ensemble | – | – | $NVT$ | $NPT$ | $NVT$ | $NVT$ |
| $\tau_\text{T}$ | – | – | 0.1/0.1/0.1 | 0.1/0.1/0.1–0.1/0.1 | 0.1/0.1–2 | 2 |
| $\tau_\text{p}$ | – | – | – | 1.0 | – | – |
| Solute constraints | None | | | All bonds | | |
| Restraint force (kJ mol$^{-1}$ nm$^{-1}$) | None | None | 1000 | 1000–100–0 | 0 | 0 |
| Fe center restraint | frozen | frozen | 99999 | 99999 | 99999 | 99999 |
| | Minimization vac. | Minimization solv. | Heating | $p$ equilibration | $T$ equilibration | Production |
| No. of steps | $F_{\text{tol}} = 1$ (200 steps) kJmol$^{-1}$nm$^{-1}$ | $F_{\text{tol}} = 1000$ kJmol$^{-1}$nm$^{-1}$ | 100000 | 200000 + 50000 +200000 | 200000 +1000000 +1000000* | 47305000 |
| Propagation | Steepest descent | Steepest descent | | Leap-frog Verlet | | |
| Ensemble | – | – | $NVT$ | $NPT$ | $NVT$ | $NVT$ |
| $\tau_\text{T}$ | – | – | 0.1/0.1/0.1 | 0.1/0.1 | 0.1/0.1–2–0.1* | 2 |
| $\tau_\text{p}$ | – | – | – | 1.0 | – | – |
| Solute constraints | None | | | All bonds | | |
| Restraint force (kJ mol$^{-1}$ nm$^{-1}$) | None | 1000 | 1000 | 1000–100–0 | 0 | 0 |
| Fe center restraint | frozen | frozen | 99999 | 99999 | 99999–50000* | 50000 |

* Run was necessary to avoid system collapse at the production stage

11

Table 6: Compositions of the simulation systems

|  | *open* conformation | *closed* conformation |
|---|---|---|
| No. of solute atoms | 10852 | 10846 |
| No. of Na$^+$/Cl$^-$ ions | 128/96 | 99/67 |
| No. of water molecules | 31800 | 20889 |
| Initial box size $(a, b/\text{nm}, V/\text{nm}^3)$ | 9.277, 12.273, 1056.34 | 8.056, 11.390, 739.201 |
| Equilibrated box $(a, b/\text{nm}, V/\text{nm}^3)$ | 9.276, 12.271, 1055.85 | 8.007, 11.321, 725.755 |

basis set [49] on all other atoms. The resolution of the identity approximation with corresponding auxiliary basis sets [50] was invoked to speed up the calculations. [51] The same force-field setup as in the MD simulations was used for the MM region. QM–MM electrostatic interactions were calculated fully (no cut-off), and the QM and MM systems were coupled with the charge-shift scheme. [39,52–54] Where the QM–MM boundary cut through a covalent bond, the QM region was saturated with a H link atom. MM partial charges of residues cut by a QM–MM boundary were corrected such that the QM and MM parts had integer charges. The correction charge (which is usually of the order of $0.01e$) was distributed equally over the entire residue involved, leading to very minor differences between the MM and QM/MM partial charges for these residues (see Tables 1, 2, 3).

Two different QM regions were defined (see Fig. 7 of the main paper). The first QM region contains the Fe center, both CO ligands, the side chain of the iron-coordinating cysteine (Cys176), the gunaylylpyridinol ligand up to the phosphate linker and the hydride-acceptor up to the phenyl part. This first region was used to study $H_2$ activation reactions at the Fe center and hydride transfer to methenyl-$H_4$MPT$^+$. The second region did not contain the substrate, but instead the side chain of His14. This region was used for the investigation of proton transfer from the pyridinol hydroxyl group to His14.

## 2 Supplementary results

### 2.1 Effect of the basis set

The $H_2$ cleavage reaction in the *closed* conformation with the snapshot at 11 ns was recalculated with a def2-TZVP basis set [48] on all atoms. Also with the larger basis, no stable hydride species could be located, and structure optimizations converged to the reduced substrate. The reaction energy was $-16.0$ kcal/mol compared to $-18.7$ kcal/mol with the smaller basis set. Hence, a larger basis set has no significant quantitative or qualitative effect on structures; the small effect on energies leaves conclusions and interpretation unaffected.

### 2.2 Potential energy surface scans

To estimate the reaction barriers for the $H_2$ cleavage reaction with proton transfer to oxypyridine (snapshot at 11 ns) and of the proton transfer from thy pyridi-

nol OH group to His14 in the *open* conformation (snapshot at 10.78 ns), we calculated energy profiles along scan coordinates that closely resemble the reaction coordinates; the profiles are plotted in Fig. 4. The scans provide an upper bound for the reaction barrier. It was not possible to locate transition states for the two reactions. It is likely that the flat profile of the two reactions caused difficulties in numerical hessian calculations.

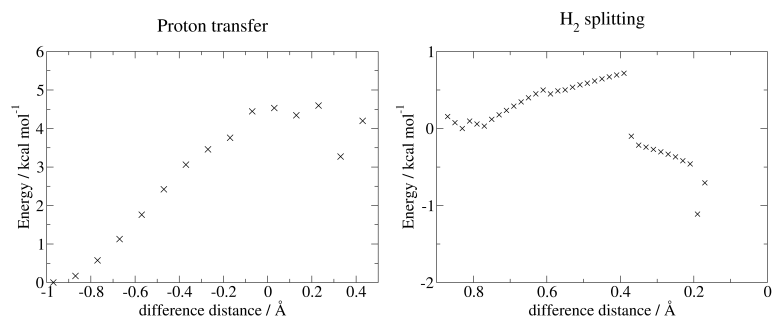

Figure 4: Plots of the potential energy surface scans. Left: Proton transfer from pyridinol OH to His14; the scan coordinate is the difference between the O–H and H–N$^\varepsilon$ bond lengths. Right: $H_2$ cleavage with proton transfer to the oxypyridine O.; the scan coordinate is the difference between the O–H and H–H bond lengths.

13

# References

[1] Duan, Y.; Wu, C.; Chowdhury, S.; Lee, M. C.; Xiong, G.; Zhang, W.; Yang, R.; Cieplak, P.; Luo, R.; Lee, T.; Caldwell, J.; Wang, J.; Kollman, P. *J. Comput. Chem.* **2003**, *24*, 1999–2012.

[2] Lee, M. C.; Duan, Y. *Proteins* **2004**, *55*, 620–634.

[3] Wang, J.; Wolf, R. M.; Caldwell, J. W.; Kollman, P. A.; Case, D. A. *J. Comput. Chem.* **2004**, *25*, 1157–1174.

[4] Frisch, M. J.; Trucks, G. W.; Schlegel, H. B.; Scuseria, G. E.; Robb, M. A.; Cheeseman, J. R.; Scalmani, G.; Barone, V.; Mennucci, B.; Petersson, G. A.; Nakatsuji, H.; Caricato, M.; Li, X.; Hratchian, H. P.; Izmaylov, A. F.; Bloino, J.; Zheng, G.; Sonnenberg, J. L.; Hada, M.; Ehara, M.; Toyota, K.; Fukuda, R.; Hasegawa, J.; Ishida, M.; Nakajima, T.; Honda, Y.; Kitao, O.; Nakai, H.; Vreven, T.; Montgomery, J. A., Jr.; Peralta, J. E.; Ogliaro, F.; Bearpark, M.; Heyd, J. J.; Brothers, E.; Kudin, K. N.; Staroverov, V. N.; Kobayashi, R.; Normand, J.; Raghavachari, K.; Rendell, A.; Burant, J. C.; Iyengar, S. S.; Tomasi, J.; Cossi, M.; Rega, N.; Millam, J. M.; Klene, M.; Knox, J. E.; Cross, J. B.; Bakken, V.; Adamo, C.; Jaramillo, J.; Gomperts, R.; Stratmann, R. E.; Yazyev, O.; Austin, A. J.; Cammi, R.; Pomelli, C.; Ochterski, J. W.; Martin, R. L.; Morokuma, K.; Zakrzewski, V. G.; Voth, G. A.; Salvador, P.; Dannenberg, J. J.; Dapprich, S.; Daniels, A. D.; Farkas, O.; Foresman, J. B.; Ortiz, J. V.; Cioslowski, J.; Fox, D. J. *Gaussian 09 Revision C.01*, Gaussian Inc. Wallingford CT 2009.

[5] Becke, A. D. *J. Chem. Phys.* **1993**, *98*, 5648–5652.

[6] Lee, C.; Yang, W.; Parr, R. G. *Phys. Rev. B* **1988**, *37*, 785–789.

[7] Stephens, P. J.; Devlin, F. J.; Chabalowski, C. F.; Frisch, M. J. *J. Phys. Chem.* **1994**, *98*, 11623–11627.

[8] Hariharan, P.; Pople, J. *Theoret. Chim. Acta* **1973**, *28*, 213–222.

[9] Clark, T.; Chandrasekhar, J.; Spitznagel, G. W.; von Rague é Schleyer, P. *J. Comput. Chem.* **1983**, *4*, 294–301.

[10] Tomasi, J.; Mennucci, B.; Cammi, R. *Chem. Rev.* **2005**, *105*, 2999–3094.

[11] Scalmani, G.; Frisch, M. J. *J. Chem. Phys.* **2010**, *132*, 114110.

[12] Ahlrichs, R.; Bär, M.; Häser, M.; Horn, H.; Kölmel, C. *Chem. Phys. Lett.* **1989**, *162*, 165–169.

[13] *TURBOMOLE V6.3 2011, a development of University of Karlsruhe and Forschungszentrum Karlsruhe GmbH, 1989-2007, TURBOMOLE GmbH, since 2007; available from* http://www.turbomole.com.

[14] Singh, U. C.; Kollman, P. A. *J. Comput. Chem.* **1984**, *5*, 129–145.

[15] Klamt, A.; Schüürmann, G. *J. Chem. Soc. Perk. Trans. 2* **1993**, 799–805.

[16] Dunning, T. H. *J. Chem. Phys.* **1989**, *90*, 1007–1023.

[17] Woon, D. E.; Dunning, T. H. *J. Chem. Phys.* **1995**, *103*, 4572–4585.

[18] Balabanov, N. B.; Peterson, K. A. *J. Chem. Phys.* **2005**, *123*, 064107.

[19] Hiromoto, T.; Warkentin, E.; Moll, J.; Ermler, U.; Shima, S. *Angew. Chem. Int. Ed.* **2009**, *48*, 6457–6460.

[20] Sousa da Silva, A.; Vranken, W. *BMC Research Notes* **2012**, *5*, 367.

[21] Wang, J.; Wang, W.; Kollman, P. A.; Case, D. A. *J. Mol. Graph. Model.* **2006**, *25*, 247–260.

[22] Case, D. A.; Darden, T. A.; III, T. E. C.; Simmerling, C. L.; Wang, J.; Duke, R. E.; Luo, R.; Walker, R. C.; Zhang, W.; Merz, K. M.; Roberts, B.; Hayik, S.; Roitberg, A.; Seabra, G.; Swails, J.; Götz, A. W.; Kolossváry, I.; Wong, K. F.; Paesani, F.; Vanicek, J.; Wolf, R. M.; Liu, J.; Wu, X.; Brozell, S. R.; Steinbrecher, T.; Gohlke, H.; Cai, Q.; Ye, X.; Wang, J.; Hsieh, M.-J.; Cui, G.; Roe, D. R.; Mathews, D. H.; Seetin, M. G.; Salomon-Ferrer, R.; Sagui, C.; Babin, V.; Luchko, T.; Gusarov, S.; Kovalenko, A.; Kollman, P. A. *AMBER13*, University of California, San Francisco 2012.

[23] Carvalho, A. T. P.; Swart, M. *J. Chem. Inf. Model.* **2014**, *54*, 613–620.

[24] Jorgensen, W. L.; Chandrasekhar, J.; Madura, J. D.; Impey, R. W.; Klein, M. L. *J. Chem. Phys.* **1983**, *79*, 926–935.

[25] Hiromoto, T.; Ataka, K.; Pilak, O.; Vogt, S.; Stagni, M. S.; Meyer-Klaucke, W.; Warkentin, E.; Thauer, R. K.; Shima, S.; Ermler, U. *FEBS Lett.* **2009**, *583*, 585–590.

[26] DeLano, W. L. *The PyMOL Molecular Graphics System, Version 1.6*, 2002.

[27] Word, J. M.; Lovell, S. C.; Richardson, J. S.; Richardson, D. C. *J. Mol. Biol.* **1999**, *285*, 1735–1747.

[28] Olsson, M. H. M.; Søndergaard, C. R.; Rostkowski, M.; Jensen, J. H. *J. Chem. Theory Comput.* **2011**, *7*, 525–537.

[29] Pilak, O.; Mamat, B.; Vogt, S.; Hagemeier, C. H.; Thauer, R. K.; Shima, S.; Vonrhein, C.; Warkentin, E.; Ermler, U. *J. Mol. Biol.* **2006**, *358*, 798–809.

[30] Pettersen, E. F.; Goddard, T. D.; Huang, C. C.; Couch, G. S.; Greenblatt, D. M.; Meng, E. C.; Ferrin, T. E. *J. Comput. Chem.* **2004**, *25*, 1605–1612.

[31] Berendsen, H. J. C.; van der Spoel, D.; van Drunen, R. *Comput. Phys. Commun.* **1995**, *91*, 43–56.

[32] Van Der Spoel, D.; Lindahl, E.; Hess, B.; Groenhof, G.; Mark, A. E.; Berendsen, H. J. C. *J. Comput. Chem.* **2005**, *26*, 1701–1718.

[33] Hess, B.; Kutzner, C.; van der Spoel, D.; Lindahl, E. *J. Chem. Theor. Comput.* **2008**, *4*, 435–447.

[34] Pronk, S.; Pll, S.; Schulz, R.; Larsson, P.; Bjelkmar, P.; Apostolov, R.; Shirts, M. R.; Smith, J. C.; Kasson, P. M.; van der Spoel, D.; Hess, B.; Lindahl, E. *Bioinformatics* **2013**, *29*, 845–854.

[35] Bussi, G.; Donadio, D.; Parrinello, M. *J. Chem. Phys.* **2007**, *126*, 014101.

[36] Senn, H. M.; Thiel, W. *Angew. Chem. Int. Ed.* **2009**, *48*, 1198–1229.

[37] Humphrey, W.; Dalke, A.; Schulten, K. *J. Mol. Graphics* **1996**, *14*, 33–38.

[38] *ChemShell, a Computational Chemistry Shell*, see www.chemshell.org.

[39] Sherwood, P.; de Vries, A. H.; Guest, M. F.; Schreckenbach, G.; Catlow, C. A.; French, S. A.; Sokol, A. A.; Bromley, S. T.; Thiel, W.; Turner, A. J.; Billeter, S.; Terstegen, F.; Thiel, S.; Kendrick, J.; Rogers, S. C.; Casci, J.; Watson, M.; King, F.; Karlsen, E.; Sjøvoll, M.; Fahmi, A.; Schäfer, A.; Lennartz, C. *Theochem* **2003**, *632*, 1–28.

[40] Metz, S.; Kästner, J.; Sokol, A. A.; Keal, T. W.; Sherwood, P. *WIRS Comput. Mol. Sci.* **2014**, *4*, 101–110.

[41] *TURBOMOLE V6.5 2013, a development of University of Karlsruhe and Forschungszentrum Karlsruhe GmbH, 1989-2007, TURBOMOLE GmbH, since 2007; available from* http://www.turbomole.com.

[42] Billeter, S. R.; Turner, A. J.; Thiel, W. *Phys. Chem. Chem. Phys.* **2000**, *2*, 2177–2186.

[43] Kästner, J.; Thiel, S.; Senn, H. M.; Sherwood, P.; Thiel, W. *J. Chem. Theor. Comput.* **2007**, *3*, 1064–1072.

[44] Perdew, J. P. *Phys. Rev. B* **1986**, *33*, 8822–8824.

[45] Becke, A. D. *Phys. Rev. A* **1988**, *38*, 3098–3010.

[46] Tao, J.; Perdew, J. P.; Staroverov, V. N.; Scuseria, G. E. *Phys. Rev. Lett.* **2003**, *91*, 146401.

[47] Grimme, S.; Antony, J.; Ehrlich, S.; Krieg, H. *J. Chem. Phys.* **2010**, *132*, 154104.

[48] Weigend, F.; Ahlrichs, R. *Phys. Chem. Chem. Phys.* **2005**, *7*, 3297–3305.

[49] Schäfer, A.; Horn, H.; Ahlrichs, R. *J. Chem. Phys.* **1992**, *97*, 2571.

[50] Weigend, F. *Phys. Chem. Chem. Phys.* **2006**, *8*, 1057–1065.

[51] Eichkorn, K.; Weigend, F.; Treutler, O.; Ahlrichs, R. *Theor. Chem. Acc.* **1997**, *97*, 119–124.

[52] Sherwood, P.; de Vries, A. H.; Collins, S. J.; Greatbanks, S. P.; Burton, N. A.; Vincent, M. A.; Hillier, I. H. *Faraday Discuss.* **1997**, *106*, 79–92.

[53] König, P. H.; Hoffmann, M.; Frauenheim, T.; Cui, Q. *J. Phys. Chem. B* **2005**, *109*, 9082–9095.

[54] Lin, H.; Truhlar, D. G. *J. Phys. Chem. A* **2005**, *109*, 3991–4004.

16



\end{document}